\documentclass[12pt]{article}
\pdfoutput=1

\usepackage{draft,hyperref,tikz,cancel}
\usepackage[nosort]{cite}

\usetikzlibrary{snakes}
\usetikzlibrary{shapes.misc}

\newcommand{\bS}{\mathbb{S}}
\newcommand{\bT}{\mathbb{T}}
\newcommand{\cU}{{\cal U}}

\begin{document}

\begin{titlepage}

\begin{center}

\hfill \\
\hfill \\
\vskip 1cm

\title{
Semiclassical OPE coefficients from 3D gravity
}

\author{Chi-Ming Chang$^\bullet$, Ying-Hsuan Lin$^\circ$}

\address{$^\bullet$Center for Theoretical Physics and Department of Physics,  \\
University of California, Berkeley, CA 94704 USA}
\address{$^\circ$Jefferson Physical Laboratory, Harvard University, \\
Cambridge, MA 02138 USA}

\email{cmchang@berkeley.edu, yhlin@physics.harvard.edu}

\end{center}

\abstract{We present a closed form expression for the semiclassical OPE coefficients that are universal for all 2D CFTs with a ``weak'' light spectrum, by taking the semiclassical limit of the fusion kernel.  We match this with a properly regularized and normalized bulk action evaluated on a geometry with three conical defects, analytically continued in the deficit angles beyond the range for which a metric with positive signature exists.  The analytically continued geometry has a codimension-one coordinate singularity surrounding the heaviest conical defect.  This singularity becomes a horizon after Wick rotating to Lorentzian signature, suggesting a connection between universality and the existence of a horizon.
} 

\vfill

\end{titlepage}

\eject

\tableofcontents

\section{Introduction}

Conformal field theories in two dimensions are constrained by modular invariance and crossing symmetry.  The two have more in common than is often appreciated: both connect the UV to the IR, and strongly constrain the defining data -- the spectrum and OPE coefficients -- of the conformal field theory \cite{Polyakov:1974gs,Ferrara:1973yt,Mack:1975jr,Belavin:1984vu,Cardy:1986ie,Knizhnik:1984nr,Gepner:1986wi,Bouwknegt:1992wg,Verlinde:1988sn,Dijkgraaf:1988tf,Moore:1988uz,Moore:1988ss,Rattazzi:2008pe,ElShowk:2012ht,Fitzpatrick:2012yx,Komargodski:2012ek,Alday:2015eya,Chang:2015aa,Lin:2015wcg,Kim:2015oca}.  In fact, under a conformal map, the torus partition function can be recast as a four-point function of $\bZ_2$ twist fields in the symmetric product orbifold theory, and modular invariance of the former under $\tau \to -{1/\tau}$ is equivalent to crossing symmetry of the latter under $x \to 1-x$ \cite{Zamolodchikov:1987ae,Knizhnik:1987zp,Dixon:1986qv,Hamidi:1986vh,Lunin:2000yv,Witten:2007kt,Headrick:2010zt,Chang:2015aa}.  A most famous consequence of modular invariance is the universal growth of the density of states at high energies, known as the Cardy formula \cite{Cardy:1986ie}, whose application to holographic contexts characterizes the growth of black hole microstates \cite{Strominger:1996sh,Strominger:1997eq}.

The first half of this paper is an application of Cardy's idea to crossing symmetry.  It is therefore instructive to first give a brief review of the Cardy formula, from Cardy's original derivation, to its application to black hole microstate counting by Strominger and Vafa \cite{Strominger:1996sh,Strominger:1997eq}, and a more careful justification of the validity of this application by Hartman, Keller, and Stoica \cite{Hartman:2014oaa}.

We begin with Cardy's derivation.  In the Hamiltonian formalism, the torus partition function is a thermal partition function, given by a sum over states in the Hilbert space of the CFT on a spatial circle, weighted by the Boltzmann factor.  
It is dominated at extreme low temperatures by the contribution of the vacuum state alone. Modular invariance equates this to a high temperature partition function, which receives contributions from states of very high energies. The Cardy formula, which characterizes the density of states at energies much higher than the vacuum Casimir energy, can be read off from the high temperature partition function by a inverse Laplace transform that takes us from the canonical to the micro-canonical ensemble.  Using standard CFT terminology, the Cardy formula describes the exponential growth of the density of states for scaling dimensions much larger than the central charge,
\ie\label{CardyV}
\Delta \gg c.
\fe
Since the derivation of the Cardy formula only relies on general axioms of 2D CFTs, it holds universally for all 2D CFTs with a unique and isolated vacuum state.  The form of the formula only depends on the central charge, or equivalently on the vacuum Casimir energy
\ie
E_0 = - {c \over 12}.
\fe

Quantum gravity on anti-de Sitter space obeys a seemingly different universality.  As long as the low energy effective theory is described by Einstein gravity, the black hole entropy follows the Bekenstein-Hawking law and is proportional to the area of the horizon.  In the seminal work by Strominger and Vafa \cite{Strominger:1996sh,Strominger:1997eq}, it was argued that for the class of black holes whose near horizon region is described by a locally AdS$_3$ geometry, the microstates can be counted by the degrees of freedom in the CFT living on the boundary of the three-dimensional bulk.  In this respect, the Bekenstein-Hawking area law and the Cardy formula are in fact two facades of one universality.

However, a remaining puzzle in this story, as pointed out in the original paper by Strominger and Vafa \cite{Strominger:1996sh} and later sharpened by Hartman, Keller, and Stoica in \cite{Hartman:2014oaa}, is that the Bekenstein-Hawking area law and Cardy's derivation are valid in different parameter regimes. 
For the area law to be valid, the bulk curvature has to be weak to suppress higher derivative corrections to Einstein gravity, which means that the AdS radius must be large in Planck units.  This bulk (semiclassical) limit translates in the CFT to a large central charge limit,
\ie
\label{SemiCardy}
c \to \infty, \quad \Delta \sim c,
\fe
in contrast to the regime of validity \eqref{CardyV} of Cardy's derivation.  We will refer to this as the semiclassical limit.  Curiously, in many supersymmetric examples where black hole microstates can be counted by certain indices \cite{Dijkgraaf:1996it,Dijkgraaf:1996xw,Maldacena:1999bp}, one sees that the index actually obeys the Cardy formula in this extended regime of validity.

This puzzle was recently resolved in \cite{Hartman:2014oaa}, where the authors showed that as long as the spectrum of the CFT satisfies a certain sparseness condition, the regime of validity of the Cardy formula can be extended to
\ie
\label{HartmanCardy}
c \to \infty, \quad h, \, \bar h \geq {c \over 12}.
\fe
The sparseness condition requires that the spectrum is sufficiently sparse in the range
\ie
h < {c \over 24} \quad \text{or} \quad \bar h < {c \over 24},
\fe
so that the partition function is dominated by the vacuum state for temperatures below the Hawking-Page phase transition \cite{Hawking:1982dh,Witten:1998aa}.  Taking the large central charge limit of the Laplace transform gives a formula for the density of states that is identical to the Cardy formula but with a different regime of validity \eqref{HartmanCardy}.

We will run a story parallel to the above in deriving universal consequences of crossing symmetry.  The four-point function of identical operators of weight $(h_{ext}, \bar h_{ext})$ has a Virasoro block decomposition
\ie\label{VirasoroBlockDecomposition}
\sum_{h, \bar h} C^2(h_{ext}, \bar h_{ext}, h, \bar h) {\cal F}(h_{ext},h,c|x) \overline{{\cal F}(h_{ext},h,c|x)},
\fe
where the expansion coefficients $C^2(h_{ext},h,\bar h_{ext},\bar h)$ are sums of the square of the OPE coefficients.  In previous work \cite{Chang:2015aa}, the present authors formulated a ``weakness'' condition which if obeyed by the ``light'' spectrum 
\ie
h < {m}_1(h_{ext}) \, c \quad \text{or} \quad \bar h < {m}_1(\bar h_{ext}) \, c,
\fe
then the OPE coefficients in the semiclassical limit \eqref{SemiCardy} follow a universal decay formula \eqref{Universal} for large enough weights (``heavy'' spectrum)
\ie
h > {m}_2(h_{ext}) \, c \quad \text{and} \quad \bar h > {m}_2(\bar h_{ext}) \, c.
\fe
$m_1$ and $m_2$ are solutions to certain equations \eqref{m1} and \eqref{m2} involving the semiclassical Virasoro block.  This is directly parallel to the universal spectrum story of \cite{Hartman:2014oaa}, and the analogy is summarized in Table~\ref{Tab:LMH}.  In fact, there is a direct connection between the two: under a conformal transformation, the torus partition function is equal to the four-point function of the $\bZ_2$ twist fields in the symmetric orbifold CFT, for which\footnote{The ground state of the $\bZ_2$-twisted sector in the symmetric product orbifold theory has weight \cite{Lunin:2000yv}
\ie
{c/2 \over 24} \left( 2 - {1 \over 2} \right) = {c \over 32},
\fe
where $c/2$ is the central charge of the single copy theory.
}
\ie
h_{ext} = \bar h_{ext} = {c\over32}, \quad m_1({c\over32}) = {1 \over 24}, \quad m_2({c\over32}) = {1 \over 12}.
\fe
After correcting for the conformal factor, the universal formula \eqref{Universal} exactly reproduces the Cardy formula \cite{Chang:2015aa}.

\begin{table}[t]
\centering
\begin{tabular}{|c|c|c|c|}
\hline
spectrum & torus & four-point & relevance
\\\hline\hline
light & $h < {c \over 24}$ or $\bar h < {c \over 24}$ & $h < {m}_1(h_{ext}) \, c$ or $\bar h < {m}_1(\bar h_{ext}) \, c$ & sparseness/weakness
\\
heavy & $h > {c \over 12}$ and $\bar h > {c \over 12}$ & $h > {m}_2(h_{ext}) \, c$ and $\bar h > {m}_2(\bar h_{ext}) \, c$ & universality
\\\hline
\end{tabular}
\caption{The light and heavy spectrum as defined in \cite{Hartman:2014oaa} in their analysis of the torus partition function, and the analogs for the four-point function.}
\label{Tab:LMH}
\end{table}

In this paper, we derive the universal formula for the OPE coefficients following a logic similar to the derivation of the Cardy formula, and deduce a closed form expression by making use of an amazing identity of Ponsot and Teschner \cite{Ponsot:1999uf,Teschner:2001rv,Ponsot:2003ju}, that relates Virasoro blocks to their image under crossing $x \to 1-x$, known as the fusion transformation.  The fusion transformation which we spell out in Section~\ref{Sec:Fusion} is expressed as a contour integral over Virasoro blocks in the cross channel, weighted by the so-called fusion kernel, which is yet another contour integral.  In the semiclassical limit, both contour integrals can be evaluated by the steepest descent method, and the universal formula is nothing but the semiclassical limit of the fusion kernel.

One important feature of the universal formula is that the OPE coefficients decay exponentially in the large dimension limit, and saturate the bound of \cite{Maldacena:2015iua}.  This is in contrast to the Cardy formula which describes an exponential growth in the density of states.  The qualitative difference is solely due to the aforementioned conformal factor. 

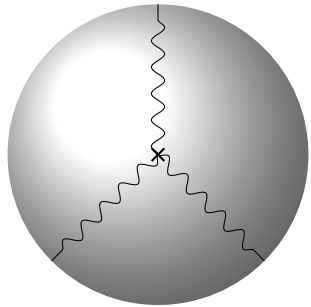
\begin{figure}
\centering
\begin{tikzpicture}
\node[circle,shading=ball,minimum width=4cm,ball color=white] (ball) at (0,0) {};
\node[draw, cross out, inner sep=2pt, thick] at (0,0) {};
\draw [snake=snake,line after snake=1mm] (0,0) -- (0,2);
\draw [snake=snake,line after snake=1mm] (0,0) -- (1.41421,-1.41421);
\draw [snake=snake,line after snake=1mm] (0,0) -- (-1.41421,-1.41421);
\end{tikzpicture}
\caption{Three conical defects joining in 3D hyperbolic space.}
\label{Fig:Conical}
\end{figure}

We will explore the gravity interpretation of the universal formula for the OPE coefficients by considering CFT operators that correspond to conical defects.  These defects have masses below the BTZ black hole threshold, which means that the scaling dimensions of their dual operators are bounded by
$\Delta < {c \over 12}$.
A natural conjecture is that the universal formula describes the cubic interaction of the conical defects in the bulk, and should be reproduced by  the regularized Einstein-Hilbert action evaluated on a geometry with three joining conical defects, as shown in Figure~\ref{Fig:Conical}.  The gravity action in 3D hyperbolic space can be rewritten as a Liouville action on the conformal boundary, and the conical defects enter as boundary conditions on the Liouville field \cite{Krasnov:2000zq,Krasnov:2000ia}.  We explicitly solve the Liouville equation (which is equivalent to solving the bulk Einstein equation), and find that by analytically continuing in the deficit angles beyond the range where a real solution exists, the properly normalized gravity action matches exactly with the semiclassical OPE coefficients of the CFT.  The analytically continued metric contains a singular surface, which can be interpreted as a horizon once we Wick rotate to Lorentzian signature.  We comment on this horizon in our discussions section.

The organization of this paper is as follows.  Section~\ref{Sec:Semi} reviews the conformal bootstrap analysis of 2D CFTs in the semiclassical limit, and explains why under a certain ``weakness'' condition the semiclassical OPE coefficients are just given by the fusion kernel.  Section~\ref{Sec:Fusion} is devoted to a careful treatment of the semiclassical limit of the fusion transformation.  Section~\ref{BulkAction} computes the gravity partition function in the presence of three conical defects, and shows how it matches with the fusion kernel.  Section~\ref{Sec:Discuss} ends with some discussions and open questions.
Appendix~\ref{Sec:Special} defines the special functions appearing in the fusion transformation and computes their semiclassical limits.  Appendix~\ref{SemiclassicalCB} discusses the convergence properties of semiclassical Virasoro blocks.  Appendix~\ref{OSLiouville} computes the on-shell classical Liouville action in the presence of three conical defects, which is used in Section~\ref{BulkAction} to compute the bulk action.  Appendix~\ref{Sec:violation} discusses subtleties in regularizing the gravity action.  Appendix~\ref{Sec:Liouville} reviews the semiclassical limit of the Liouville CFT.

To illustrate the central idea of this paper, let us begin with a simple exercise using just scaling blocks.  This exercise was considered in \cite{Pappadopulo:2012jk,Maloney}.

\subsection{An exercise with scaling blocks}
\label{Sec:Scaling}

A four-point function can be written as a sum over intermediate states in a particular channel
\ie
\langle \phi_a(x_1) \phi_b(x_2) \phi_c(x_3) \phi_d(x_4) \rangle
= \sum_i
\begin{tikzpicture}[baseline={([yshift=-.5ex]current bounding box.center)},vertex/.style={anchor=base,
    circle,fill=black!25,minimum size=18pt,inner sep=2pt}]
\draw (-.5,0) -- (.5,0);
\node at (0, 0.25) {$\phi_i$};
\draw (-.5,0) -- (-1,.5);
\node at (-1.5, 1) {$\phi_a(x_1)$};
\draw (-.5,0) -- (-1,-.5);
\node at (-1.5, -1) {$\phi_b(x_2)$};
\draw (.5,0) -- (1,.5);
\node at (1.5, 1) {$\phi_c(x_3)$};
\draw (.5,0) -- (1,-.5);
\node at (1.5, -1) {$\phi_d(x_4)$};
\end{tikzpicture}
\fe
Each state $\phi_i$ contributes a term proportional to the scaling block $x^{\Delta_i - \Delta_a - \Delta_b}$ (borrowing terminology from \cite{Kim:2015oca}).  If we assume unitarity, then in the limit of $(\phi_a, \phi_b)$ and $(\phi_c, \phi_d)$ being pairwise close, the operator $\phi_0$ that has the low scaling dimension dominates the sum.  Up to a conformal factor, the four-point function is well-approximated by
\ie
x^{\Delta_{0} - \Delta_a - \Delta_b} + {\cal O}(x^{\Delta_1 - \Delta_a - \Delta_b}),
\fe
where
$
x = {(x_1-x_2) (x_3-x_4) \over (x_1-x_4) (x_3-x_2)}
$
is the cross ratio, and $\Delta_0$ and $\Delta_1$ are the lowest and second lowest scaling dimensions that appear in this channel.  When all four external operators are identical, $\phi_0$ is simply the identity operator.

In the cross channel
\ie
\langle \phi_a(x_1) \phi_b(x_2) \phi_c(x_3) \phi_d(x_4) \rangle
= \sum_i
\begin{tikzpicture}[baseline={([yshift=-.5ex]current bounding box.center)},vertex/.style={anchor=base,
    circle,fill=black!25,minimum size=18pt,inner sep=2pt}]
\draw (0,-.5) -- (0,.5);
\node at (-.25,0) {$\phi_i$};
\draw (0,.5) -- (-.5,1);
\node at (-1, 1.5) {$\phi_a(x_1)$};
\draw (0,-.5) -- (-.5,-1);
\node at (-1, -1.5) {$\phi_b(x_2)$};
\draw (0,.5) -- (.5,1);
\node at (1,1.5) {$\phi_c(x_3)$};
\draw (0,-.5) -- (.5,-1);
\node at (1,-1.5) {$\phi_d(x_4)$};
\end{tikzpicture}
\fe
the four-point function in the limit of $x \to 0$ has a binomial expansion
\ie
& x^{\Delta_{0} - \Delta_a - \Delta_b} + {\cal O}(x^{\Delta_{1} - \Delta_a - \Delta_b})
\\
& = \sum_{n=0}^\infty \left\{
\begin{pmatrix}
\Delta_{0} - \Delta_a - \Delta_b
\\
n
\end{pmatrix} + \# 
\begin{pmatrix}
\Delta_{1} - \Delta_a - \Delta_b
\\
n
\end{pmatrix}
+ \dotsb
\right\}
(x-1)^n.
\fe
The corrections are suppressed when $n$ is large, hence the coefficients in the scaling block decomposition of the four-point function follow a binomial distribution
\ie
(-1)^n \begin{pmatrix}
\Delta_{0} - \Delta_a - \Delta_b
\\
n
\end{pmatrix}
\sim { n^{\Delta_a + \Delta_b - \Delta_0 - 1} \over \Gamma( \Delta_a + \Delta_b - \Delta_0 ) }.
\fe
When all external operators are identical, the contribution of an operator of weight $\Delta_\phi$ to the four-point function with $x = {1\over2}$ is
\ie
{ \Delta_\phi^{2\Delta_a - 1} \over \Gamma(2\Delta_a) } \times \left( 1\over2 \right)^{\Delta_\phi - 2\Delta_a} \left[ 1 + {\cal O} \left( \Delta_a \over \Delta_\phi \right) \right], \quad \Delta_\phi \equiv 2\Delta_a + n,
\fe
which for large enough $\Delta_\phi$ satisfies the general bootstrap bound obtained in \cite{Kim:2015oca}.

\section{Semiclassical OPE coefficients}
\label{Sec:Semi}

In previous work \cite{Chang:2015aa}, by analyzing crossing symmetry in the semiclassical limit, the present authors derived a ``weakness'' condition under which the OPE coefficients must follow a universal formula, that is expressed as the difference of two semiclassical Virasoro blocks.  
Section~\ref{Sec:Bootstrap} reviews this bootstrap analysis.
Section~\ref{Sec:Recast} then draws an analogy between the the Cardy formula and the universal formula for the OPE coefficients, and explains that the universal formula is nothing but the fusion kernel, which is the kernel of an integral transformation that relates Virasoro blocks to their images under crossing.

\subsection{A universal formula from conformal bootstrap}
\label{Sec:Bootstrap}

Given a family of CFTs labeled by increasing and unbounded central charge $c$, the semiclassical limit is the limit of $c \to \infty$ while simultaneously scaling the operator weights with $c$.  A more careful definition is given in \cite{Chang:2015aa}.

In \cite{Chang:2015aa}, we analyzed the semiclassical limit of the crossing equation for identical external operators.  Let us review this analysis.  In \cite{Zamolodchikov:426555}, it was observed that the Virasoro block ``exponentiates'' in the semiclassical limit,
\ie
\label{BlockClassExp}
{\cal F}(h_{ext}, h, c | x) = \exp\left[-{c\over 6} f\Big({h_{ext}\over c},{h\over c}\Big|x\Big) + {\cal O}(c^0) \right].
\fe
In Appendix \ref{SemiclassicalCB}, we examine the validity of this formula in more details.  The function $f$ will be referred as the ``semiclassical Virasoro block". Our main interest is the coefficient $C^2(h_{ext}, \bar h_{ext}, h, \bar h)$ in the Virasoro block decomposition \eqref{VirasoroBlockDecomposition} of the four-point function, which is the OPE coefficient squared smeared over operators with weights lying in a small range around $(h, \bar h)$.  Let us make two remarks:
\begin{enumerate}  
\item  In order to satisfy the crossing equation, $C^2(h_{ext}, \bar h_{ext}, h, \bar h)$ can at most grow exponentially with the central charge.
\item  At a generic cross ratio, the four-point function should be dominated by a single term in the Virasoro block decomposition in either channel. 
\end{enumerate} 
If the CFT has a vacuum state and a ``weak''  light spectrum (defined below), then for cross ratios within the interval $(0, {1\over2})$, the dominant term in one channel is the vacuum block, and the coefficient $C^2(h_{ext}, \bar h_{ext}, h, \bar h)$ of the dominant term in the other channel is given by the bootstrap equation.

Explicitly, the weakness condition requires that
\ie\label{weakness}
C^2(h_{ext}, \bar h_{ext}, h, \bar h) < \exp\left\{{c\over 6}\left[ f\Big({h_{ext}\over c},{h\over c}\Big|{1\over 2}\Big) -  f\Big({h_{ext}\over c},0\Big|{1\over 2}\Big)\right] + (\text{anti-holo})\right\}
\fe
in the ``light'' spectrum range
\ie
h < {m}_1(h_{ext}) \, c \quad \text{or} \quad \bar h < {m}_1(\bar h_{ext}) \, c.
\fe
When this is satisfied, by varying the cross ratio inside $(0, {1\over2})$, we find that the OPE coefficients in the ``heavy'' spectrum range
\ie
h \geq {m}_2(h_{ext}) \, c \quad \text{and} \quad \bar h \geq {m}_2(\bar h_{ext}) \, c
\fe
obey a universal formula
\ie
\label{Universal}
& C^2(h_{ext}, \bar h_{ext}, h, \bar h)
\\
& = \exp\left\{{c\over 6}\left[ f\Big({h_{ext}\over c},{h\over c}\Big|1-\widehat x(h)\Big) -  f\Big({h_{ext}\over c},0\Big|\widehat x(h)\Big)\right] + (\text{anti-holo}) + {\cal O}(\log c) \right\},
\fe
where $\widehat x(h)$ is the solution to
\ie
\label{xhat}
{d \over dx} f\Big({h_{ext}\over c},{h\over c}\Big|x\Big)\Big|_{x = 1-\widehat x(h)} + {d \over dx}f\Big({h_{ext}\over c},0\Big|x\Big)\Big|_{x = \widehat x(h)} = 0.
\fe

The functions ${m}_1(h_{ext})$ and ${m}_2(h_{ext})$ that define the ranges of the light and heavy spectrum are solutions to the equations
\ie
\label{m1}
{d \over dx} f\Big({h_{ext}\over c},m_1\Big|x\Big)\Big|_{x = {1\over2}} = 0
\fe
and
\ie
\label{m2}
{d \over dx}  f\Big({h_{ext}\over c},m_2\Big|x\Big)\Big|_{x = {1\over2}} + {d \over dx} f\Big({h_{ext}\over c},0\Big|x\Big)\Big|_{x = {1\over2}} =0.
\fe
The values and properties of $m_1$ and $m_2$ are the subjects of \cite{Chang:2015aa}.  Qualitatively, when $h_{ext} \ll c$, we have
\ie
m_1 c \approx {\sqrt2 h_{ext}}, \quad m_2 c \approx {2\sqrt2 h_{ext}},
\fe
and as $h_{ext}$ increases, the ratios $m_1 c / h_{ext}$ and $m_2 c / 2 h_{ext}$ decrease monotonically, but never go below one.

\subsection{Recasting as fusion kernel}
\label{Sec:Recast}

We now present the universal formula in a way that is more physically illuminating.  The logic here will be analogous to Cardy's derivation of the universal growth of the density of states.  However, the weakness condition and the value of $m_1$ must still come from the conformal bootstrap analysis.

The assumption of a weak light spectrum is equivalent to the requirement that in the semiclassical limit, the vacuum block dominates a four-point function with identical external operators, for cross ratios in the entire interval $(0, {1\over2})$.  When this happens, the crossing equation to all perturbative orders in $1/c$ is equivalent to the fusion transformation \cite{Ponsot:1999uf,Teschner:2001rv,Ponsot:2003ju} of the vacuum block,
\ie
\label{FusionIntegral}
{\cal F}(h_{\A_{ext}}, 0, c | x) = \int_{\bS} d\A_t\, {\bf F}^{(c)}_{0, \A_t}[\A_{ext}]\,
{\cal F}(h_{\A_{ext}},h_{\A_t}, c | 1- x),
\fe
where
\ie\label{LiouvillehcQ}
h_{\A} = \A(Q-\A), \quad c = 1+6Q^2, \quad Q = b + 1/b.
\fe
The fusion kernel\footnote{In the notation of \cite{Ponsot:1999uf,Teschner:2001rv,Ponsot:2003ju}, it is ${\bf F}_{\A_s, \A_t}
\!\! \begin{bmatrix}\A_{ext} & \A_{ext} \\ \A_{ext} & \A_{ext}\end{bmatrix}$.} ${\bf F}^{(c)}_{\A_s \A_t}[\A_{ext}]$ has a contour integral expression \eqref{FusionMatrix} that involves some special functions $\Gamma_b, S_b$.  These functions are reviewed in Appendix~\ref{Sec:Special}.  The contour $\bS$ runs from ${Q \over 2}$ to ${Q\over2} + i\infty$, but picks up residues of certain poles, the details of which are spelled out in Section~\ref{Sec:Fusion}.

Up to a Jacobian factor
\ie
{d\A_t \over dh_{\A_t}} = {1 \over \sqrt{4h_{\A} - Q^2}}, 
\fe
the right hand side of \eqref{FusionIntegral} is essentially the decomposition of the vacuum block in the cross channel.  If we assume that this integral is dominated near a particular $\A_t$ in the semiclassical limit, then we immediately realize that the holomorphic part of the universal formula for the OPE coefficients is equal to the fusion kernel, perturbatively to all orders in $1/c$.  By varying the cross ratio $x$ inside $(0, {1\over2})$, this equivalence holds for all weights $h_{\A_t}$ greater than $m_2(h_{ext}) \, c$.

Including also the anti-holomorphic part, we conclude that under the weakness condition, the OPE coefficients obey a universal formula
\ie
\label{CFusion}
C^2(h_{\A_{ext}}, \bar h_{\A_{ext}}, h_{\A}, \bar h_{\A}) = {{\bf F}^{(c)}_{0, \A_t}[\A_{ext}]\,
 \over \sqrt{4h_{\A} - Q^2}} \times (\text{anti-holo}) \left[ 1 + {\cal O}(e^{-\#c}) \right]
\fe
for large enough weights
\ie
h_{\A_t} \geq m_2(h_{ext}) \, c, \quad \bar h_{\A_t} \geq m_2(\bar h_{ext}) \, c.
\fe
The steepest descent approximation of the fusion kernel in the semiclassical limit will be the subject of Section~\ref{Sec:Fusion}.  Two comments are in order:
\begin{enumerate}
\item  The above analysis can be generalized to two pairs of external operators, or some appropriate average of operators, as long as the vacuum block appears in one channel.
\item  Perturbatively to all orders in $1/c$, the universal formula for the OPE coefficients is completely factorized into a holomorphic and an anti-holomorphic piece.  This is simply because the vacuum block is factorized.
\end{enumerate}

\section{Semiclassical limit of the fusion transformation}

\label{Sec:Fusion}

This section is devoted to a careful treatment of the semiclassical limit of the fusion transformation.  The special functions that appear here 
are defined and their properties reviewed in Appendix~\ref{Sec:Special}.

The fusion transformation relates a Virasoro block to Virasoro blocks in the cross channel through the following expression \cite{Ponsot:1999uf,Teschner:2001rv,Ponsot:2003ju}:
\ie\label{FusionTrans}
{\cal F}(h_{\A_{ext}}, h_{\A_s}, c | x) = \int_{\mathbb S} d\A_t\, {\bf F}^{(c)}_{\A_s, \A_t}[\A_{ext}]\,
 {\cal F}(h_{\A_{ext}},h_{\A_t}, c | 1- x).
\fe
For simplicity we specialize to the case of identical external operators with weight $h_{ext}$, and only consider real $x$.  The variables $\A$, $Q$ and $b$ are related to the weight $h_{\A}$ and central charge $c$ by \eqref{LiouvillehcQ} and $\A_{ext}$ takes value in the physical region $[0,{Q\over2}] \cup {Q\over2}+i\bR_{\geq0}$, such that $h_{\A_{ext}} = \A_{ext} (Q - \A_{ext})$ is real and non-negative.  Since we are interested in the large $c$ limit, we will assume that $b$ is positive.  The contour $\bS$ runs from $Q\over2$ to ${Q\over2} + i\infty$ while circumventing poles in the fusion kernel in a manner that will be prescribed below.

The fusion kernel ${\bf F}^{(c)}_{\A_s,\A_t}[\A_{ext}]$ has a contour integral representation 
\ie
\label{FusionMatrix}
{\bf F}^{(c)}_{\A_s,\A_t} [\A_{ext}]\,
 &= P_b(\A_s, \A_t, \A_{ext}) \times {1\over i} \int_{\bT}ds \, T_b(\A_s, \A_t, \A_{ext}, s),
\fe
where $P_b$ and $T_b$ are
\ie
\label{FusionMatrixTerms}
& P_b(\A_s, \A_t, \A_{ext}) = {\Gamma_b(2Q - 2\A_{ext} - \A_t)\Gamma_b(\A_t)^2\Gamma_b(Q-\A_t)^2 \over \Gamma_b(2Q-2\A_{ext}-\A_s)\Gamma_b(\A_s)^2\Gamma_b(Q-\A_s)^2}
\\
& \times { \Gamma_b(Q-2\A_{ext}+\A_t) \Gamma_b(2\A_{ext}+\A_t-Q)\Gamma_b(2\A_{ext}-\A_t) \over \Gamma_b(Q-2\A_{ext}+\A_s) \Gamma_b(2\A_{ext}+\A_s-Q)\Gamma_b(2\A_{ext}-\A_s)} \times {\Gamma_b(2Q-2\A_s)\Gamma_b(2\A_s)\over \Gamma_b(Q-2\A_t)\Gamma_b(2\A_t-Q)},
\\
& T_b(\A_s, \A_t, \A_{ext}, s) = {S_b(U_1+s)S_b(U_2+s)S_b(U_3+s)S_b(U_4+s)\over S_b(V_1+s)S_b(V_2+s)S_b(V_3+s)S_b(V_4+s)},
\\
&U_1=\A_s,\quad U_2=Q+\A_s-2\A_{ext},\quad U_3=\A_s+2\A_{ext}-Q,\quad U_4=\A_s,
\\
&V_1=Q+\A_s-\A_t,\quad V_2=\A_s+\A_t,\quad V_3=2\A_s,\quad V_4=Q.
\fe
$\Gamma_b(x)$ is a meromorphic function that has poles at $x=-mb-n/b$ for non-negative integers $m$ and $n$, and
\ie
S_b(x) \equiv {\Gamma_b(x)\over \Gamma_b(Q-x)}.
\fe
See Appendix~\ref{Sec:Special} for a definition of these special functions.

\subsection{Fusion kernel}
\label{Sec:FusionKernel}

The integrand $T_b$ as a function of $s$ has poles at 
\ie
&{\cal U}_1:~~s=-\A_s-mb-n/b,~~~m,n=0,1,\cdots,
\\
&{\cal U}_2:~~s=-\A_s+2\A_{ext}-mb-n/b,~~~m,n=1,2,\cdots,
\\
&{\cal U}_3:~~s=-\A_s-2\A_{ext}-mb-n/b,~~~m,n=-1,0,1,\cdots,
\\
&{\cal U}_4:~~s=-\A_s-mb-n/b,~~~m,n=0,1,\cdots,
\\
&{\cal V}_1:~~s=-\A_s+\A_t+mb+n/b,~~~m,n=0,1,\cdots,
\\
&{\cal V}_2:~~s=-\A_s-\A_t+mb+n/b,~~~m,n=1,2,\cdots,
\\
&{\cal V}_3:~~s=-2\A_s+mb+n/b,~~~m,n=1,2,\cdots.
\\
&{\cal V}_4:~~s=mb+n/b,~~~m,n=0,1,\cdots.
\fe
When some of these arrays of poles overlap, we turn on small imaginary regulators
\ie
\label{Regulators}
\A_{ext} \to \A_{ext} + i \epsilon_{ext}, \quad \A_s \to \A_s + i \epsilon_s, \quad \epsilon_{ext}, \, \epsilon_s > 0
\fe
to separate the poles.  Along the contour $\bS$, the imaginary part of $\A_t$ is always positive. The contour $\bT$ of the $s$-integral in \eqref{FusionMatrix} runs from $-i\infty$ to $i\infty$ such that the poles $\bigcup_i {\cal U}_i$ lie to the left of the contour, and the poles $\bigcup_i {\cal V}_i$ lie to the right.

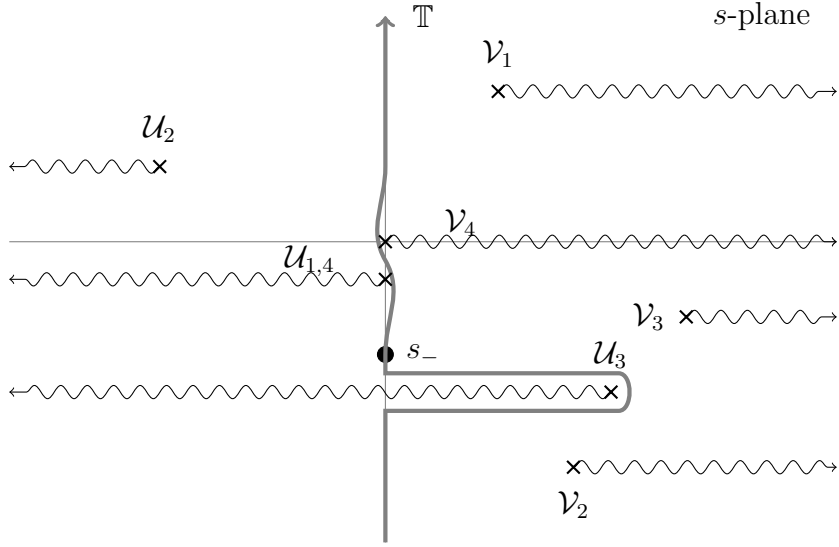
\begin{figure}[t]
\centering
\begin{tikzpicture}
\node at (5, 3) {$s$-plane};
\filldraw (0,-1.5) circle (3pt);
\node at (.5,-1.5) {$s_-$};
\draw [->, color=gray] (0,-4) -- (0,3);
\draw [->, color=gray] (-5,0) -- (6,0);
\draw [->, color=gray, ultra thick] (0,-4) -- (0,-2.25) -- (3.1,-2.25) to[out=0,in=0] (3.1,-1.75) -- (0,-1.75) to[out=90,in=-90] (0,-1.5) to[out=90,in=-60] (0,-.25) to[out=120, in=-90] (0, 1) -- (0, 3);
\node at (.5,3) {$\bT$};
\draw [->,snake=snake, line after snake=1mm] (0,-.5) -- (-5,-.5);
\node[draw, cross out, inner sep=2pt, thick] at (0,-.5) {};
\node at (-1,-.25) {$\cU_{1,4}$};
\draw [->,snake=snake, line after snake=1mm] (-3,1) -- (-5,1);
\node[draw, cross out, inner sep=2pt, thick] at (-3,1) {};
\node at (-3,1.5) {$\cU_2$};
\draw [->,snake=snake, line after snake=1mm] (3,-2) -- (-5,-2);
\node[draw, cross out, inner sep=2pt, thick] at (3,-2) {};
\node at (3,-1.5) {$\cU_3$};
\draw [->,snake=snake,line after snake=1mm] (1.5,2) -- (6,2);
\node[draw, cross out, inner sep=2pt, thick] at (1.5,2) {};
\node at (1.5,2.5) {$\cV_1$};
\draw [->,snake=snake, line after snake=1mm] (2.5,-3) -- (6,-3);
\node[draw, cross out, inner sep=2pt, thick] at (2.5,-3) {};
\node at (2.5,-3.5) {$\cV_2$};
\draw [->,snake=snake, line after snake=1mm] (4,-1) -- (6,-1);
\node[draw, cross out, inner sep=2pt, thick] at (4,-1) {};
\node at (3.5,-1) {$\cV_3$};
\draw [->,snake=snake, line after snake=1mm] (0,0) -- (6,0);
\node[draw, cross out, inner sep=2pt, thick] at (0,0) {};
\node at (1,.25) {$\cV_4$};
\end{tikzpicture}
\caption{The contour $\mathbb T$ in the definition of the fusion kernel.  When 
$\eta_{ext}$ and $\eta_s$ are real, small and positive regulators are turned on so that the arrays of poles do not overlap, see \eqref{Regulators}.  This figure is drawn with the choice $2\epsilon_{ext} : \epsilon_s : \text{Im}\,\eta_t = 3 : 1 : 5$.  The solid dot is the dominant critical point $s_-$.
}
\label{Fig:ContourC}
\end{figure}

We are after the semiclassical limit of the fusion kernel with $\A_s = 0$, which is the limit of\footnote{This is the same as the semiclassical limit defined in Section~\ref{Sec:Bootstrap}, $c \to \infty$ with $h/c$ fixed.
}
\ie\label{SemiLimit}
b \to 0, \quad \eta_{ext} \equiv b \A_{ext}~~\text{and}~~\eta_{t} \equiv b \A_{t} ~~ \text{fixed}.
\fe
In this limit, the arrays of poles of $T_b$ create branch cuts which the contour $\bT$ must circumvent, see Figure~\ref{Fig:ContourC}.\footnote{Before taking the semiclassical limit, the contour $\bT$ can freely pass through the zeros of $T_b$.  After taking the limit, these zeros create branch cuts which the $\bT$ should also circumvent.}  The semiclassical limit of the special functions $\Gamma_b$ and $S_b$ are computed in Appendix~\ref{Sec:Special}.  The result is, loosely speaking,
\ie\label{SemiPT}
b^2\log\Gamma_b(y/b) \to G(y), \quad b^2\log S_b(y/b) \to H(y),
\fe
where the functions $G$ and $H$ are defined as
\ie
G(y) \equiv - \int_{1/2}^y \log\Gamma(z) dz, \quad H(y) \equiv G(y) - G(1-y).
\fe
$G$ has a branch cut on the negative real axis; we define $G(y)$ to be real when $y$ is on the positive real axis (since there $\Gamma_b(y/b)$ is real and positive), and by analytic continuation on the rest of $\bC \setminus \bR_{<0}$.  


We perform a steepest descent approximation to the contour integral of $T_b$ over $s$.  To simplify the analysis, let us assume that $\eta_{ext}, \eta_t \in (0, {1\over2})$.  In the semiclassical limit, the exponent of $T_b$ becomes
\ie\label{SemiTbIG}
\lim_{b\to 0} b^2 \log T_b(\eta_s/b, \eta_t/b, \eta_{ext}, s/b) = \sum_{i=1}^4 H(u_i + s) - H(v_i + s), ~~ u_i \equiv b U_i, ~~ v_i \equiv b V_i,
\fe
with $U_i$ and $V_i$ defined in \eqref{FusionMatrixTerms}.  The steepest descent equation is\footnote{The value of $\log \sin$ is chosen such that the identity
\ie
\log\pi - \log\sin\pi y = \log\Gamma(y) + \log\Gamma(1-y)
\fe 
is satisfied, and $\log\Gamma$ is defined such that it is real on the positive real axis and only has a branch cut along the negative real axis.
}
\ie
2\pi i N = \sum_{i = 1}^4 \log \sin\pi (u_i + s) - \sum_{i = 1}^4 \log \sin\pi (v_i + s),
\fe
where $N$ labels the sheet.  This equation is invariant under $s \to s+1$ shifts, so let us focus on the strip $-{1\over2} < s \leq {1\over2}$.  When $\eta_s = i \epsilon_s$ where $\epsilon_s$ is a positive small regulator as in \eqref{Regulators}, there is one critical point $s_-$ lying on the negative imaginary axis of sheet $N = 0$, and another $s_+$ lying on the positive imaginary axis of a different sheet $N = -1$, and their distances to the origin are both of order $\epsilon_s$.\footnote{With $t \equiv s/\eta_s$ fixed, taking exponential of the steepest descent equation gives
\ie
\left( {\sin\pi\eta_t \over \sin(2\pi\eta_{ext})} \right)^2 = { (1+t)^2 \over t(2+t) } + {\cal O}(\eta_s^2),
\fe
which has two solutions.  One of them has zero imaginary part before taking exponential, while the other has $-2\pi i$.
}

We presently argue that, in the $\epsilon_s \to 0$ limit, $T_b(s_-) \to T_b(0)$ is a dominant contribution to the contour integral.  Firstly, $s_-$ lies on the contour $\bT$, and one can check that $\text{Re}\,\log T_b(s_-)$ is smaller than the maximum\footnote{The maximum occurs at the point which is the lift of $s_+$ to the original sheet.} of $\text{Re}\,\log T_b(\bT)$ by an amount of order ${\cal O}(\epsilon_s)$.  Following \cite{Witten:2010aa}, we define gradient flows generated by the real part of \eqref{SemiTbIG} as a Morse function. Whether a critical point $s_i$ contributes to the contour integral depends on whether the upwards (with increasing real part) gradient flow line out of the critical point intersects the contour $\bT$.  For the ones that do,
\ie
\label{FlowLine}
\text{Re}\,\log T_b(s_i) < \text{Re}\,\log T_b(\bT) < \text{Re}\,\log T_b(s_-) + {\cal O}(\epsilon_s),
\fe
and therefore $T_b(s_i)$ will be less dominant than $T_b(s_-)$.  The only exception is $s_+$, which is ${\cal O}(\epsilon_s)$ distance away from $s_-$.  Nonetheless, even if $s_+$ contributes, its contribution is of the same order as $s_-$ in the $\eta_s \to 0$ limit, $T_b(s_+) \to T_b(0)$.
To conclude, in the semiclassical limit, the contour integral is approximated by
\ie\label{SemiTb}
&\lim_{b\to 0} b^2  \log T_b(0, \eta_t/b, \eta_{ext}, 0) 
\\
&= - 2(G(0) - G(1)) + [ G(2\eta_{ext}-1) - G(2\eta_{ext}) ] + [ G(1 - 2\eta_{ext}) - G(2 - 2\eta_{ext}) ].
\fe
The semiclassical limit of the prefactor $P_b$ is straightforwardly computed to be
\ie
\label{SemiPb}
&\lim_{b\to 0} b^2  \log P_b(0, \eta_t / b, \eta_{ext} / b) 
\\
&= - F(2\eta_{ext} + \eta_t - 1) - 2 F(\eta_t) + F(2\eta_{ext} - 1) - F(2\eta_{ext} - \eta_t) + F(2\eta_{ext}) 
\\
& \hspace{.5in} - G(1-2\eta_t) - G(2\eta_t-1) 
+ 1 - 2G(1),
\fe
with $F$ defined as
\ie
F(y) \equiv - G(y) - G(1-y).
\fe
Combining the two, we arrive at the semiclassical fusion kernel for the vacuum block,
\ie\label{SemiF}
&\lim_{b\to 0} b^2  \log {\bf F}^{(6/b^2)}_{0, \eta_t/b} [\eta_{ext}/b]\,
\\
& = - F(2\eta_{ext} + \eta_t - 1) - F(2\eta_{ext} - \eta_t) - 2 F(\eta_t) - 2G(2\eta_{ext})  - 2G(2-2\eta_{ext})
\\
& \hspace{.5in} - G(1-2\eta_t) - G(2\eta_t-1) 
+ 1 - 2G(0).
\fe
If all $\eta_{ext}$ and $\eta_t$ are real, then this expression may sit on a branch cut.
However, this does not happen because $\text{Im}\,\eta_t$ is assumed to be positive, in accordance with the prescription of the contour $\bS$.

\subsection{Fusion transformation}
\label{Sec:FusionTrans}

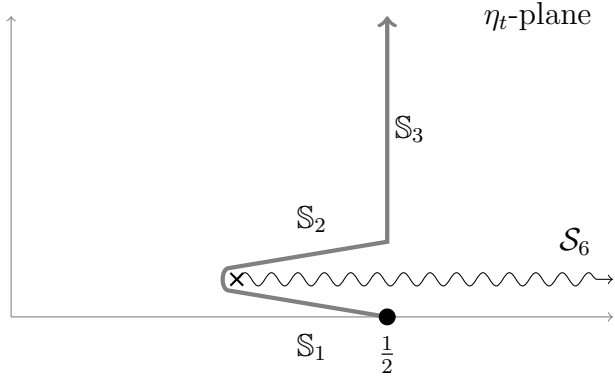
\begin{figure}\label{Scontour}
\centering
\begin{tikzpicture}

\node at (2, 4) {$\eta_t$-plane};
\draw [->, color=gray] (-5,0) -- (-5,4);
\draw [->, color=gray] (-5,0) -- (3,0);

\draw [->,snake=snake,line after snake=1mm] (-2,.5) -- (3,.5);
\node[draw, cross out, inner sep=2pt, thick] at (-2,.5) {};

\node at (2.5, 1) {${\cal S}_6$};

\draw [->, color=gray, ultra thick] (0,0) -- (-2.1,0.35) to[out=180,in=180] (-2.1,.65) -- (0,1) -- (0,4);

\node at (-1, -.4) {$\bS_1$};
\node at (-1, 1.3) {$\bS_2$};
\node at (.3, 2.5) {$\bS_3$};

\node at (0, -0.5) {${1\over 2}$};

\filldraw (0,0) circle (3pt);

\end{tikzpicture}
\caption{The contour $\bS$ when $\eta_{ext} \leq {1\over4}$.}
\label{Fig:ContourS}
\end{figure}

Let us proceed to evaluating the semiclassical limit of the $\A_t$-integral in the fusion transformation \eqref{FusionTrans} with another steepest descent approximation.  We first analyze the pole structure of the integrand, which consists of the fusion kernel and the cross-channel Virasoro block, and give a prescription of the contour $\mathbb S$.  We then show that the critical point(s) must lie on $(0, {1 \over 2}) \cup {1 \over 2} + i\bR_{\geq0}$, and that the fusion kernel at the critical point(s) is real.

\paragraph{Prescription of contour $\mathbb S$}

Recall that $\Gamma_b(x)$ has poles at $x=-mb-n/b$ for non-negative integers $m$ and $n$. Thus $P_b(0, \A_t, \A_{ext})$ as a function of $\A_t$ has poles at 
\ie
\label{PolesP}
&{\cal S}_1:~~\A_t=-2\A_{ext}+mb+n/b,~~~m,n=2,3,\cdots,
\\
&{\cal S}_2:~~\A_t=-mb-n/b,~~~m,n=0,1,\cdots,
\\
&{\cal S}_3:~~\A_t=mb+n/b,~~~m,n=1,2,\cdots,
\\
&{\cal S}_4:~~\A_t=2\A_{ext}-mb-n/b,~~~m,n=1,2,\cdots,
\\
&{\cal S}_5:~~\A_t=-2\A_{ext}-mb-n/b,~~~m,n=-1,0,1,\cdots,
\\
&{\cal S}_6:~~\A_t=2\A_{ext}+mb+n/b,~~~m,n=0,1,\cdots.
\fe
The Virasoro block as a function of $\A_t$ has poles when the dimension $h_{\A_t}$ of the internal operator becomes degenerate,
\ie
\label{PolesV}
&{\cal S}_7:~~\A_t={1\over 2}(mb+n/b),~~~m,n=2,3,\cdots,
\\
&{\cal S}_8:~~\A_t=-{1\over 2}(mb+n/b),~~~m,n=0,1,2,\cdots.
\fe

When $\A_{ext} \in({Q\over4},{Q\over 2})  \cup {Q\over2}+i\bR_{\geq0}$, the contour $\mathbb S$ can simply be chosen to run along the line ${Q\over2}+i\bR_{\geq0}$, since all the poles are away from this contour.  But when $\A_{ext} \in (0, {Q\over4}]$, the poles ${\cal S}_5$ and ${\cal S}_6$ cross the imaginary axis.  We recall from the previous subsection the regularization $\A_{ext} \to \A_{ext}+ {i\epsilon_{ext} / b}$. The poles ${\cal S}_5$ are on the lower half plane, and the poles ${\cal S}_6$ are on the upper half plane.  The contour is deformed such that it circumvents the poles ${\cal S}_6$, as shown in Figure~\ref{Fig:ContourS}.

\paragraph{Steepest descent approximation of the $\eta_t$-integral}

As in the case of the fusion kernel, the poles \eqref{PolesP} and \eqref{PolesV} accumulate into branch cuts in the semiclassical limit.

Let us first consider the case $\eta_{ext} \in (0,{1\over4}]$.  As shown in Figure~\ref{Fig:ContourS}, we split the contour into three pieces,
\ie
&{\mathbb S}_1:~~\eta_t = {1 \over 2} \to 2(\eta_{ext}+ i\epsilon_{ext}),
\\
&{\mathbb S}_2:~~\eta_t = 2(\eta_{ext}+ i\epsilon_{ext}) \to {1 \over 2} + i \epsilon_t,
\\
&{\mathbb S}_3:~~\eta_t = {1 \over 2} + i \epsilon_t \to {1 \over 2} + i \infty,
\fe
where $\epsilon_t > 2\epsilon_{ext}$ is a small regulator.

Along the contour ${\mathbb S}_3$, the semiclassical fusion kernel \eqref{SemiF} is manifestly real, and so is the the semiclassical Virasoro block since we assumed that $x$ is real from the beginning.  Hence the exponent of the integrand, given by the sum of the two, is also real along ${\mathbb S}_3$.  Therefore, this contour coincides with a gradient flow line generated by a Morse function defined as the real part of this exponent.  By the same argument as we gave near \eqref{FlowLine}, the steepest descent approximation of this integral can only receive dominant contribution from either critical points that lie on this contour, or from the boundary point $\eta_t=0$.  The critical points are the solutions to the equation\footnote{This equation is solved by the same $\eta_t$ that solves the bootstrap equation of \cite{Chang:2015aa},
\ie
f'(\eta_{ext}, 0 | x) + f'(\eta_{ext}, \eta_t | 1-x) = 0,
\fe
where $f'$ denotes the derivative with respect to $x$.  To see this, let $\eta_t(x)$ denote the critical point.  Take derivative with respect to $x$ on the semiclassical fusion transformation
\ie
f(\eta_{ext}, 0 | x) = f(\eta_{ext}, \eta_t(x) | 1-x) - \lim_{b\to0} b^2 \log {\bf F}_{0\eta_t(x)/b}[\eta_{ext}/b],
\fe
and reorganize into
\ie
\hspace{-.25in} f'(\eta_{ext}, 0 | x) + f'(\eta_{ext}, \eta_t(x) | 1-x) = {d\eta_t(x) \over dx} {d \over d\eta_t} \left[ f(\eta_{ext}, \eta_t(x) | 1-x) - \lim_{b\to0} b^2 \log {\bf F}_{0,\eta_t(x)/b}[\eta_{ext}/b] \right].
\fe
}
\ie\label{SPE3}
0 &=  - \log\gamma(2\eta_{ext}+\eta_t-1) + \log\gamma(2\eta_{ext}-\eta_t) - 2 \log\gamma(\eta_t) - 2\log\Gamma(1-2\eta_t)
\\
& \hspace{.5in} + 2\log\Gamma(2\eta_t-1)
 - {d \over d\eta_t} f(\eta_{ext}, \eta_t | 1-x).
\fe
Let us stress again that while there may be other solutions to \eqref{SPE3} that do not lie on the contour $\bS_3$, those critical points do not contribute to the integral, or are less dominant.

Along the contour ${\mathbb S}_2$, by use of the recursion relations \eqref{GIHPeriods} for $G$ and $F$, the semiclassical fusion kernel \eqref{SemiF} can be rewritten in a manifestly real form: 
\ie\label{FKR2}
&\lim_{b\to 0} b^2  \log {\bf F}^{(6/b^2)}_{0 ,\eta_t/b}[\eta_{ext}/b]
\\
&=G(2\eta_{ext} + \eta_t )   + G(2-2\eta_{ext} - \eta_t ) + G(2\eta_{ext} - \eta_t +1 )+ G(1-2\eta_{ext} + \eta_t)
\\
& \hspace{-.25in} + 2(\eta_t - 2\eta_{ext})- 2G(2\eta_{ext})  - 2G(2-2\eta_{ext}) + F(2\eta_t) - 2 F(\eta_t) + 1 + F(0)
\\
& \hspace{-.5in} + (2\eta_{ext} + \eta_t - 1) \log (1 - 2\eta_{ext} - \eta_t  ) + (2\eta_{ext} - \eta_t )  \log (\eta_t  - 2\eta_{ext} ) - (2\eta_t-1) \log (1-2\eta_t).
\fe
Since the semiclassical Virasoro block is also real ($x$ is real), the dominant critical point(s) must lie on the contour $\bS_2$.  The steepest descent equation is
\ie
\label{SPE1}
&\hspace{-.5in}  - \log\Gamma(2\eta_{ext}+\eta_t) + \log\Gamma(2-2\eta_{ext}-\eta_t) + \log\Gamma(2\eta_{ext}-\eta_t+1) - \log\Gamma(1-2\eta_{ext}+\eta_t) 
\\
&\hspace{0in} + 2\log\gamma(2\eta_t) -2 \log\gamma(\eta_t)+ \log (1 - 2\eta_{ext} - \eta_t  ) -  \log (\eta_t  - 2\eta_{ext} )  - 2 \log (1-2\eta_t) 
\\
&\hspace{.5in} - {d \over d\eta_t} f(\eta_{ext}, \eta_t | 1-x) = 0.
\fe

Finally, the exponent of the integrand along ${\mathbb S}_1$ has the same real part as the exponent along ${\mathbb S}_2$, while the imaginary part is equal to $2\pi i(2\eta_{ext}-\eta_t)$.  Hence the integral along ${\mathbb S}_1$ is bounded above by the integral along ${\mathbb S}_2$.  As far as extracting the leading exponent of the fusion transformation is concerned, we need not consider the integral along ${\mathbb S}_1$.

Now let us consider the case of $\eta_{ext} \in ({1\over4},{1\over2}) \cup {1\over2}+i\bR_{\geq0}$.  As noted earlier, here the contour can be chosen to be along ${1\over2}+i\bR_{\ge 0}$ since this choice does not cross any branch cut.  The semiclassical fusion kernel \eqref{SemiF} is real along this contour, so the dominant critical point(s) must lie on the contour and satisfy the steepest descent equation \eqref{SPE3}.


In summary, the fusion transformation of the vacuum block in the semiclassical limit is dominated by a Virasoro block with weight $h_t = \A_t (Q - \A_t)$. $\A_t$ lies on either $(2\A_{ext}, {Q\over2})$ or ${Q\over2} + i\bR_{\geq0}$ as a solution to one of the steepest descent equations, \eqref{SPE1} or \eqref{SPE3}, and the semiclassical fusion kernel is given in manifestly real forms by \eqref{SemiF} or \eqref{FKR2}, respectively.

We numerically verified that the semiclassical limit of the fusion kernel obtained in this section is indeed equal to the ratio between the vacuum Virasoro block and the dominant Virasoro block in the cross channel.

%

\section{Bulk action}
\label{BulkAction}

In previous sections, we argued that the OPE coefficients of 2D CFTs follow a universal formula, provided that a ``weakness'' condition is satisfied.

We propose that the universal formula can be reproduced by an analytic continuation of the regularized Einstein-Hilbert action evaluated on a geometry of three conical defects that join at a single point in the bulk.  At the boundary point of each conical defect with deficit angle $4\pi i \eta$ sits a heavy CFT operator of scaling dimension
\ie\label{DelforConical}
\Delta = h + \bar h = { c \eta(1-\eta) \over 3 }.
\fe
Throughout this section we set the AdS radius to one,
\ie
R_{AdS} = 1,
\fe
so that the central charge is related to the bulk gravitational constant by
\ie
c={3\over2G}.
\fe

In Section \ref{Probelimit}, we test our proposal in the limit of small deficit angles $\eta \ll 1$, where the conical defects can be produced by geodesic worldlines of ``heavy'' particles (a notion that we make precise later), and the on-shell Einstein-Hilbert action reduces to a worldline action.  In Section \ref{ConicalDefect}, we write down a metric that describes conical defect geometries with finite deficit angles, and compute the regularized Einstein-Hilbert action.  In both cases, we find that after an analytic continuation and proper normalization, the gravity calculation matches with the semiclassical OPE coefficients in the CFT.

\subsection{Heavy particles}
\label{Probelimit}

A ``heavy'' particle in AdS$_3$ is defined to be a particle whose mass $M$ is proportional to the Planck mass $1/G$ as we take $G \to 0$, but $G M$ is parametrically small.  In the CFT language, a heavy particle corresponds to an operator with scaling dimension $\Delta$ that scales with the central charge $c$ as we take 
$c \to \infty$, but the ratio $\Delta/c$ is parametrically small.  In both cases, it is crucial that we take the semiclassical limit before we take the small mass/scaling dimension limit.  In this limit, the relation between the mass $M$ and the scaling dimension $\Delta$ is simply
\ie
\Delta = 1 + \sqrt{1 + M^2} \to M.
\fe
Classically, the insertion of such an operator sources the worldline of a heavy particle in the bulk.

Consider a heavy particle decay process in the Poincar\'{e} patch of AdS$_3$,
\ie
ds^2 = {dy^2 + dz d\bar z \over y^2}.
\fe
A heavy scalar particle of mass $\Delta_1$ enters the AdS$_3$ at a boundary point $z_1$, and moves along a geodesic until it reaches a bulk point $\bf x$, then decays into two heavy scalar particles of masses $\Delta_2$ and $\Delta_3$.  The two particles move along their geodesics until they exit the AdS$_3$ at boundary points $z_2$ and $z_3$.  The worldline action for this decay process is\footnote{We assume that the coupling constant $\lambda$ of the bulk scalar field scales as $\lambda \sim c^\#$, and hence contribute to sub-leading $\log c$ order in the worldline action.  In large $N$ theories, the three-point coupling of single-trace operators scale as $1/N$.
}
\ie
S = \Delta_1L({\bf x},z_1)+\Delta_2L({\bf x},z_2)+\Delta_3L({\bf x},z_3),
\fe
where $L({\bf x},z')$ is the geodesic distance between a bulk point ${\bf x}=(z,\bar z,y)$ and a boundary point $(z', \bar z')$ in AdS$_3$,
\ie
L({\bf x},z')=\log \left[ y^2+|z-z'|^2 \over y \right].
\fe
With $z_1, z_2, z_3$ fixed, the bulk point $\bf x$ is chosen to minimize the worldline action.  The exponential of this action $e^{-S}$ corresponds holographically to the three-point function of
the dual scalar operators in the CFT.

The minimization problem has a solution when the triangle inequalities for $\Delta_1, \Delta_2, \Delta_3$ are obeyed, and the result is given by
\ie\label{WorldAction}
&S =  {1\over 2}(\Delta_1+\Delta_2-\Delta_3)\log|z_1-z_2|^2 + (\text{2 permutations}) - {\cal P}(\Delta_1,\Delta_2,\Delta_3),
\\
& {\cal P}(\Delta_1,\Delta_2,\Delta_3) = {1\over 2}\Delta_1\log\left[{ (\Delta_1+\Delta_2-\Delta_3)(\Delta_1+\Delta_3-\Delta_2)\over \Delta_2+\Delta_3-\Delta_1}\right] + (\text{2 permutations})
\\
& \hspace{2in} +{1\over 2} \left( \textstyle \sum_i \Delta_i \right) \left( \log{\textstyle \sum_i \Delta_i} - \log 4 \right) -  {\textstyle \sum_i \Delta_i \log \Delta_i}.
\fe
While the position dependence of the worldline action (the first term plus the two permutations) is fixed by conformal invariance, the exponential of the last term $e^{\cal P}$ should correspond holographically to the OPE coefficients in the CFT.

We would like to compare this result with the formula \eqref{CFusion} from the bootstrap analysis.  Let us set $\Delta_1=\Delta$ and $\Delta_2=\Delta_3=\Delta_{ext}$.  The semiclassical fusion kernel \eqref{FKR2} to linear order in $h_{ext}$ and $h$, combined with the anti-holomorphic part (assuming that all operators are scalars) gives\footnote{The square root is taken because we should compare the worldline action with $( C^2(h_{ext}, \bar h_{ext}, h, \bar h) )^{1\over2}$.}
\ie\label{LinearFusionK}
\log \sqrt{{\bf F} {\bf \bar F} } = {1\over 2} \Delta_{ext} \left[ (r+2) \log (r+2) - (r-2) \log (r-2) - (r+2) \log 4 \right] + {\cal O}(c^0,h_{ext}^2,h^2),
\fe
where $r = {\Delta / \Delta_{ext}}$.  An analysis of the steepest descent equation \eqref{SPE1} shows that the critical point is bounded by
\ie
\Delta > 2m_2({h_{ext}})\, c = 2\sqrt2 \Delta_{ext}.
\fe

\begin{figure}
\centering
\begin{tikzpicture}

\draw [->] (0,0)--(11,0);
\node at (11.5, 0) {$\Delta$};
\draw (0,-.1)--(0,.1);
\node at (0, -.5) {$0$};
\node at (2.6, .5) {worldline};
\node at (1.8, -.5) {weakness};
\node at (1.8, -1) {condition};

\draw [->, thick] (4.85,.5)--(5.35,.5);
\node at (8.2, .5) {analytic continuation};

\draw (3.6,-.2)--(3.6,0);
\node at (3.6, -.5) {$\sqrt{2}\Delta_{ext}$};

\draw (5.1,0)--(5.1,.2);
\node at (5.1, -.5) {$2\Delta_{ext}$};
\draw (7.2,-.2)--(7.2,0);
\node at (7.2, -.5) {$2\sqrt2\Delta_{ext}$};
\node at (9.3, -.5) {universality};

\end{tikzpicture}
\caption{Regimes of validity of the heavy particle worldline computation and the conformal bootstrap analysis.}
\label{Fig:Validity}
\end{figure}
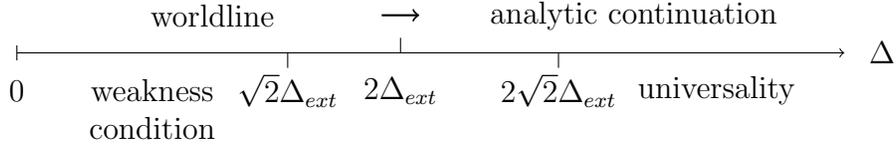

The worldline action \eqref{WorldAction} gives an almost identical formula
\ie\label{UFPP}
{\cal P}(\Delta, \Delta_{ext}, \Delta_{ext}) = {1\over 2}\Delta_{ext} \left[ (r+2) \log (r+2) + (2-r) \log (2-r) - (r+2) \log 4 \right],
\fe
except that this formula is valid for $\Delta < 2\Delta_{ext}$ since we need to obey the triangle inequality.  Using the expression \eqref{LinearSemiBlock} for the semiclassical Virasoro block to linear order in weights but exact in the cross ratio, we find that the weakness condition \eqref{weakness} is satisfied,
\ie\label{LinearWeakness}
{\cal P}(\Delta,\Delta_{ext},\Delta_{ext})\le \Delta\log\left({3+2\sqrt{2}\over 4}\right) \quad {\rm for} \quad \Delta < 2m_1(h_{ext})\, c = \sqrt{2} \Delta_{ext}.
\fe
To further compare with the fusion kernel \eqref{LinearFusionK}, we need to extend the result of the worldline computation to the region $\Delta > 2\Delta_{ext}$.  A na\"{\i}ve analytic continuation of \eqref{UFPP} could produce an ambiguous imaginary part due to the branch cut of the logarithm.  At the end of Section~\ref{ConicalDefect}, we will argue that the correct continuation does not produce any imaginary part, and hence we have an exact match.  See Figure~\ref{Fig:Validity} for a diagram depicting the different regimes of $\Delta$. 

\subsection{Conical defects}

\label{ConicalDefect}

When the boundary operator insertions have large scaling dimensions, they correspond in the bulk to objects with large masses, the back reaction can no longer be ignored.  To compute the three-point interaction in this case, we need to find a metric that descibes a hyperbolic geometry with three conical defects. See Figure \ref{Fig:Conical}.  An ansatz is
\ie\label{conicalmetric}
ds^2={4\over (1-r^2)^2}\left[dr^2+r^2e^{\varphi(z,\bar z)}dzd\bar z\right].
\fe
The coordinates $z$ and $\bar z$ are the stereographic coordinates of a two-sphere, and the whole space is topologically a three-dimensional ball with possible conical defects extending from the origin to the boundary along the radial direction at fixed angular coordinates.  The vacuum Einstein equation on this ansatz becomes the Liouville equation,\footnote{The origin of the Liouville equation is in contrast to \cite{Krasnov:2000zq,Krasnov:2000ia}. There, the Liouville equation arises from a constant negative curvature condition on the induced metric on a cutoff surface near the conformal boundary.
}
\ie
\label{LiouvilleEq}
\partial\bar\partial\varphi=2\pi \mu b^2 e^\varphi,
\fe
with the cosmological constant\footnote{Notice that $\mu$ is {\it not} the cosmological constant in the bulk gravity.} set to
$
\mu=-{1 \over 4\pi b^2}.
$

The solution for pure Euclidean AdS$_3$ is given by
\ie
e^{\varphi(z,\bar z)}= {4 \over (1+|z|^2)^2}.
\fe
We insert conical defects by introducing the boundary conditions
\ie\label{LiouvilleBc}
\varphi(z,\bar z) \to
\begin{cases}
-2\log|z|^2 & z \to \infty
\\
-2\eta_i\log|z-z_i|^2 & z\to z_i,
\end{cases}
\fe
which imply that the conical defects are scalars.\footnote{To describe conical defects with nonzero spin, one may want to consider a more general set of boundary conditions:
\ie
\varphi(z,\bar z) \to-2\eta_i\log(z-z_i)-2\bar\eta_i\log(\bar z-\bar z_i) \quad z\to z_i.
\fe
However, the single-valuedness of the metric requires $\eta_i - \bar\eta_i\in{1\over 2}\bZ$ and $\sum_i(\eta_i - \bar\eta_i)\in\bZ$, which cannot be satisfied for $\eta_i \in (0, {1\over2})$ (deficit angle less than $2\pi$). }
On the complex $z$-plane, in the small neighborhood around $z_i$, the angular part of the metric \eqref{conicalmetric} can be put into flat form 
$
dwd\bar w \propto |z-z_i|^{-4\eta_i} dzd\bar z  
$
by a multivalued coordinate transformation from $z$ to
$
w = (z-z_i)^{1-2\eta_i}.
$
The coordinate $w$ is subject to a further identification 
$
w\sim w\exp2\pi i(1-2\eta_i)
$
that creates a deficit angle $4\pi i \eta_i$ along the radial line at a fixed $z_i$ direction.

Next, we derive an expression for the on-shell gravity action for conical defect geometries.  The Einstein-Hilbert action evaluated on a space of constant curvature is given by the volume of the space,
\ie
- {1\over 16\pi G}\int d^3 x\sqrt{g}({\cal R}+2) = {1\over 4\pi G}\int d^3 x\sqrt{g} = {1\over 4\pi G}V.
\fe
Because the metric \eqref{conicalmetric} diverges as we approach the boundary $r \to 1$, the volume is also divergent.  To regularize this divergence, we introduce a cutoff surface
\ie
r=r_{max}(z,\bar z,\epsilon)<1
\fe
that approaches the boundary as the regulator $\epsilon$ is sent to zero. The regularized volume $V_\epsilon$, defined as the volume of the space inside the cutoff surface, diverges quadratically as the regulator $\epsilon$ is taken to zero. This divergence can be canceled by a boundary term on a cutoff surface
\ie
- {1\over 8\pi G}\int d^2 x\sqrt{\gamma}(K-1) = - {1\over 8\pi G}A_\epsilon,
\fe
where $A_\epsilon$ is the area. There remains a logarithmic divergence related to the Weyl anomaly of the boundary CFT. The on-shell action is given by subtracting off the logarithmic divergence and taking the regulator $\epsilon$ to zero,
\ie\label{RegAct}
S={1\over 4\pi G}\lim_{\epsilon\to 0}\left\{V_\epsilon-{1\over 2}A_\epsilon-2\pi\left[2-{\textstyle\sum_i}\eta_i(1-\eta_i)\right]\log\epsilon\right\}.
\fe
Among the terms multiplying the logarithmic divergence, the first term is from the Weyl anomaly of the Riemann sphere itself \cite{Henningson:1998gx,Krasnov:2000zq,Yin:2007at}, and the second is the Weyl anomaly of the operators.

Since our goal is to compare the on-shell gravity action with a CFT correlation function defined on the complex plane (flat), it is convenient to choose a cutoff surface whose induced metric is flat in the $\epsilon\to 0$ limit. Consider the cutoff surface,
\ie\label{CutoffSurface}
&r_{max}=1-\epsilon e^{{\varphi\over2}},
\fe
which has a flat induced metric to leading order in the $\epsilon$-expansion,
\ie
ds^2=\left({1\over \epsilon^2}-{e^{\varphi\over2}\over \epsilon}-{e^\varphi\over 4}\right)dzd\bar z +{1\over 4}(\partial \varphi dz+\bar\partial \varphi d\bar z)^2 + {\cal O}(\epsilon).
\fe
This cutoff surface approaches the origin of the unit ball when the coordinate $z$ approaches a conical defect,
\ie
1-r\propto{\epsilon \over |z-z_i|^{2\eta_i}}, \quad z\to z_i.
\fe
The $z$-integral must be constrained by $|z-z_i|>\epsilon_i$ so that the radial coordinate $r$ is always positive. This also regularizes the conical singularities.
Finally there is another divergence at $z \to \infty$, as the cutoff surface approaches the boundary. 
To regularize this divergence, we restrict the integration domain of the $z$-integral to be within $|z| \le R$.  The final $z$-integration domain is
\ie
&\Gamma=\{|z-z_i|\ge \epsilon_i, \, |z|\le R\}.
\fe

The volume and area inside the region $\Gamma$ are given by\footnote{The integration measure is $d^2z=dxdy$, $z=x+iy$.}
\ie
&V_\epsilon=\int_{\Gamma} d^2 z\left[{1\over 2\epsilon^2}-{e^{\varphi\over 2}\over 2\epsilon}+{1\over 8}e^{\varphi}\left(1+2\varphi+4\log{\epsilon\over 2}\right)\right]+{\cal O}(\epsilon), 
\\
&A_\epsilon=\int_{\Gamma} d^2 z\left[{1\over \epsilon^2}-{e^{\varphi\over 2}\over \epsilon}+{1\over 8}\left(-2 e^{\varphi}+4\partial\varphi\bar\partial\varphi\right)\right]+{\cal O}(\epsilon).
\fe
The regularized gravity action is\footnote{In the first and second equality of \eqref{V-A}, we used the Liouville equation \eqref{LiouvilleEq} and the divergence theorem.
}
\ie\label{V-A}
&V_\epsilon-{1\over 2}A_\epsilon
\\
&=\int_{\Gamma} d^2 z\left[{1\over 4}\left(\partial\varphi\bar\partial\varphi - e^{\varphi}\right) -\left(1+\log{\epsilon\over 2}\right)\partial \bar\partial \varphi- {1\over 2}\bar\partial(\varphi\partial\varphi)\right]
\\
&=\int_{\Gamma} d^2 z\left[ {1\over 4}\left(\partial\varphi\bar\partial\varphi - e^{\varphi}\right) +2\pi\left(1-{\textstyle\sum_i\eta_i}\right)\left(1+\log{\epsilon\over 2}\right)+\pi\left(\varphi_\infty-{\textstyle\sum_i\eta_i\varphi_i}\right) \right]
\\
&=\pi S_L\big|_{\pi\mu b^2=-{1\over 4}}+2\pi\left(1-\log 2 + \log\epsilon\right) (1-{\textstyle\sum_i}\eta_i)-2\pi\log R
+ 2\pi{\textstyle\sum_i}\eta_i^2\log\epsilon_i,
\fe
where $S_L$ is the classical Liouville action \cite{Zamolodchikov:1995aa}
\ie
\label{LiouvilleA}
S_L=\int_{\Gamma} d^2 z\,{1\over 4\pi}\left(\partial\varphi\bar\partial\varphi +4\pi \mu b^2 e^{\varphi}\right) +\left(\varphi_\infty+2\log R\right)-{\textstyle\sum_i}\left(\eta_i\varphi_i+2\eta_i^2\log\epsilon_i\right),
\fe
and $\varphi_i$ are defined as
\ie
\varphi_i={i\over 4\pi\eta_i}\oint_{|z|=\epsilon_i} dz \,\varphi\partial \varphi, \quad \varphi_\infty={i\over 4\pi}\oint_{|z|=R} dz \,\varphi\partial \varphi.
\fe
After subtracting off the Weyl anomalies, we end up with
\ie
S={1\over 4G}\Big[ S_L\Big|_{\pi\mu b^2=-{1\over 4}}+2\left(1-\log 2\right) \Big(1-{\textstyle\sum_i}\eta_i\Big)-2\log {(R\epsilon)}+ 2{\textstyle\sum_i}\eta_i^2\log{(\epsilon_i/\epsilon)}\Big].
\fe
The remaining task is to compute the on-shell Liouville action $S_L$.

Let us consider the case of three conical defects at $z_1, z_2, z_3$.  The solution to the Liouville equation \eqref{LiouvilleEq} with boundary conditions \eqref{LiouvilleBc} and the on-shell Liouville action are given in \cite{Zamolodchikov:1995aa,Harlow:2011aa}, which we review in Appendix \ref{OSLiouville}.  Borrowing their result, we find that if the three deficit angles satisfy the triangle inequalities and if $\sum_i \eta_i < 1$ (sum of the deficit angles is less than $4\pi$), then the gravity action is
\ie
& \hspace{-.25in} S={1\over 4G}\Big[(\delta_1+\delta_2-\delta_3)\log|z_1-z_2|^2 + (\text{2 permutations})
\Big] - {\cal P}'(\eta_1,\eta_2,\eta_3)
\\
& \hspace{-.25in} {\cal P}'(\eta_1,\eta_2,\eta_3)={1\over 4G}\Big[F(2\eta_1) - F(\eta_2+\eta_3-\eta_1) + (\text{2 permutations}) +F(0)-F({\textstyle\sum_i\eta_i})
\\
& -2\left(1-{\textstyle\sum_i\eta_i}\right)\log\left(1-{\textstyle\sum_i\eta_i}\right)
+ 2\pi i N (1-{\textstyle\sum_i\eta_i})  -2\log (R \epsilon) + 2{\textstyle\sum_i}\eta_i^2\log{(\epsilon_i / \epsilon)}\Big],
\fe
where $\delta_i=\eta_i(1-\eta_i)$ and $N \in \bZ$ labels the ambiguity in shifting the classical solution $\varphi$ by $2\pi i$.  The exponential of the on-shell Einstein-Hilbert action $e^{-S}$ has the interpretation of a three-point function, but to compare with the CFT we should consider the properly normalized version
\ie\label{ConicalOPE}
&{\cal P}(\eta_1,\eta_2,\eta_3)={\cal P}'(\eta_1,\eta_2,\eta_3)-{1\over 2} \sum_i {\cal P}'(\eta_i,\eta_i,0) + {1\over 2}{\cal P}'(0,0,0)
\\
&={c\over 6}\Big[F(2\eta_1)-F(\eta_2+\eta_3-\eta_1) +\left(1-2\eta_1\right)\log\left(1-2\eta_1\right) + (\text{2 permutations}) 
\\
&\hspace{.5in} + F(0) - F({\textstyle\sum_i\eta_i})-2\left(1-{\textstyle\sum_i\eta_i}\right)\log\left(1-{\textstyle\sum_i\eta_i}\right) \Big].
\fe 
Note that ${\cal P}(\eta,\eta,0) = 0$, and all the dependences on the regulators $R, \epsilon, \epsilon_i$ and the shift ambiguity $N$ cancel out.

Let us compare this to the bootstrap result of Section~\ref{Sec:Semi} by setting $\eta_1=\eta_t$ and $\eta_2=\eta_3=\eta_{ext}$.  The CFT operator dual to a conical defect has weight
$
h_i = \bar h_i = {c\eta_i (1-\eta_i) \over 6}.
$
One can numerically check that \eqref{ConicalOPE} interpreted as the OPE coefficients of operators dual to conical defects satisfies the weakness condition \eqref{weakness}.
To match with the bootstrap formula \eqref{CFusion}, which is given by the semiclassical fusion kernel $\bf F$ times the anti-holomorphic part, we should analytically continue \eqref{ConicalOPE} in $\eta_t$ to the triangle inequality-violating region
$
2\eta_{ext} < \eta_t \le {1\over 2}.
$
The real part of the analytically continued expression reproduces $\log \sqrt{{\bf F} {\bf \bar F}}$, where ${\bf F}$ is given in \eqref{FKR2},
 but it also contains a nonzero imaginary part 
\ie
\label{ImaginaryPart}
{\rm sgn} [ {\rm Im}(\eta_t-2\eta_{ext}) ] \, i\pi (\eta_t-2\eta_{ext}),
\fe
which comes out of the recursion relations \eqref{GIHPeriods} for $G$ and $F$.


When the triangle inequality is violated, $\eta_t > 2\eta_{ext}$, $e^\varphi$ is negative and hence the metric \eqref{conicalmetric} has indefinite signature (as can be seen from the explicit solution of $\varphi$ in Appendix~\ref{OSLiouville}).  But since the metric is still real, the volume $V$ and area $A$ should still be real.  How come the action has an imaginary part?  The answer is a failure of our current regularization scheme.  When $\eta_t > 2\eta_{ext}$, the solution of $e^{\varphi\over 2}$ has a nontrivial phase,
\ie
{ e^{\varphi \over 2} \over |e^{\varphi \over 2}| } \to {\rm sgn} [ {\rm Im}(\eta_t-2\eta_{ext}) ] \times\begin{cases}
i  & z \to z_1
\\
-i & z \to z_2, \, z_3, \, \infty,
\end{cases}
\fe
and the cutoff surface \eqref{CutoffSurface} for real $\epsilon$ becomes ill-defined.  To fix this, $\epsilon$ should be redefined with a phase to cancel the phase of $e^{\varphi \over 2}$ and make $r_{max} < 1$.  The contribution of this phase to the regularized gravity action \eqref{V-A} kills the previous imaginary part \eqref{ImaginaryPart}, and makes the answer real.

The fact that $e^\varphi$ is everywhere real and that $e^{\varphi \over 2}$ has opposite phases near $z_1$ and near $z_2, z_3, \infty$ implies that $\varphi$ has a branch cut on the $z$-plane on which $e^{\varphi \over 2}$ diverges; away from the branch cut, the phase is piecewise constant.  We regularize this divergence by cutting out a thin shell containing the branch cut from the $z$-integration domain $\Gamma$. This way the phase jump does not contribute to the classical Liouville action, and we obtain an exact match between the gravity action \eqref{ConicalOPE} and the universal formula for the OPE coefficient in the CFT.  More details of this regularization are in Appendix \ref{Sec:violation}.


%

\section{Discussions}
\label{Sec:Discuss}

In this paper, we derived a universal formula for the OPE coefficients in 2D CFTs in the semiclassical limit.  In this limit, the crossing equation is equivalent to the fusion transformation of the vacuum Virasoro block, and the universal formula for the OPE coefficients is given by the semiclassical fusion kernel.  

On the gravity side, we computed the regularized Einstein-Hilbert action in the presence of three conical defects.  At first sight, the gravity computation and the universal formula are valid in different regimes of the deficit angles.  But after an analytic continuation, the properly regularized and normalized gravity action matches exactly with the universal formula.  One peculiar feature of this analytic continuation is that the the signature of the metric becomes indefinite: the signature of the radial direction remains positive, but the signatures in the angular directions become negative.  
The CFT metric has the opposite sign compared to the induced bulk metric on the conformal boundary, 
but a sign flip 
can be achieved by an imaginary dilatation, under which the OPE coefficients are unchanged.  

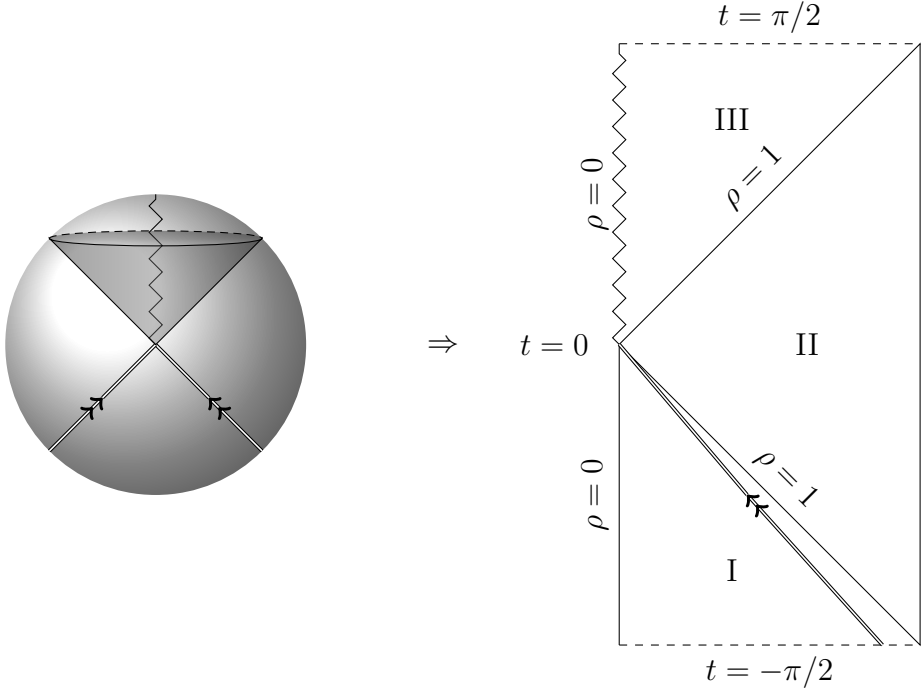
\begin{figure}[th]
\begin{center}
\begin{minipage}{.4\textwidth}
\begin{tikzpicture}
\node[circle,shading=ball,minimum width=4cm,ball color=white] (ball) at (0,0) {};
\draw [decorate,decoration=zigzag] (0,0) -- (0,2);
\draw [->>,double] (1.41421,-1.41421) -- (.707105,-.707105);
\draw [double] (.707105,-.707105) -- (0,0);
\draw [->>,double] (-1.41421,-1.41421) -- (-.707105,-.707105);
\draw [double] (-.707105,-.707105) -- (0,0);

\fill[top color=gray!50!black,bottom color=gray!10,middle color=gray,shading=axis,opacity=0.25] (0,1.41421) circle (1.41421cm and 0.1cm);
\fill[left color=gray!50!black,right color=gray!50!black,middle color=gray!50,shading=axis,opacity=0.25] (1.41421,1.41421) -- (0,0) -- (-1.41421,1.41421) arc (180:360:1.41421cm and 0.1cm);
\draw (-1.41421,1.41421) arc (180:360:1.41421cm and 0.1 cm) -- (0,0) -- cycle;
\draw[densely dashed] (-1.41421,1.41421) arc (180:0:1.41421cm and 0.1 cm);

\end{tikzpicture}
\end{minipage}
\hspace{-.5in} $\Rightarrow$ ~
\begin{minipage}{.4\textwidth}
\begin{tikzpicture}
\node at (1.5,3) {III};
\node at (2.5,0) {II};
\node at (1.5,-3) {I};

\node (I) at ( 4,0) {};
\path (I) 
	+(90:4)  coordinate (Itop)
	+(-90:4) coordinate (Ibot)
	+(180:4) coordinate (Ileft)
	;
       
\draw  (Ileft) -- (Itop) 
	node[midway,above, sloped] {$\rho = 1$}
	-- (Ibot)
	-- (Ileft) node[midway, above, sloped] {$\rho = 1$}
	-- cycle;

\draw[decorate,dashed] (0,4) 
	-- (Itop)
	node[midway,above] {$t={\pi/2}$}
	;

\draw[decorate,dashed] (0,-4)
	-- (Ibot)
	node[midway,below] {$t=-{\pi/2}$}
	;
      
\draw[decorate] (0,-4) 
	-- (0,0)
	node[midway,above,sloped] {$\rho=0$}
	;
	
\draw[decorate,decoration=zigzag] (0,0)
	node[left, inner sep=4mm] {$t=0$}
	-- (0,4)
	node[midway,above,sloped] {$\rho=0$}
	;
	
\draw[->>,decorate,double] (3.5,-4) -- (1.7,-2);
\draw[decorate,double] (1.7,-2) -- (0,0);
      
\end{tikzpicture}
\end{minipage}
\end{center}
\caption{{\bf Left}:  Two heavy particles (double-line) joining with a conical defect (zigzag), when the triangle inequality is violated.  The geometry has positive signature in the radial direction, but negative signature in the angular directions.  The cone depicts a coordinate singularity.  {\bf Right:}  After Wick rotating to Lorentzian signature, the Penrose diagram for the creation of a conical defect by two heavy particles.  Each point on this diagram away from $\rho = 0$ represents a circle, and the two particles come in from $\theta = 0, \pi$.  The coordinate singularity becomes a horizon at $\rho = 1$.  The geometry near the horizon in patch III is an FLRW universe \eqref{FLRW}, which does not see the singularity at $\rho = 0$.}
\label{Fig:Penrose}
\end{figure}

Another feature of the analytically continued metric is the existence of a codimension-one singular surface that surrounds the heaviest conical defect.  
It is a coordinate singularity and the curvature there is finite.  The metric near this singularity is
\ie
ds^2 &= {4 \over (1-r^2)^2} \left[ dr^2 -4r^2 {d\rho^2+\rho^2 d\theta^2\over (1-\rho^2)^2} \right],
\fe
and the singular surface is located at $\rho = 1$.  This metric can be rewritten in the Friedmann-Lema\^{i}tre-Robertson-Walker (FLRW) coordinates by a change of variables $r=\tanh{\tau\over 2}$.\footnote{We thank Alexander Maloney, Gim Seng Ng, and Simon F. Ross for pointing this out.}
We can further Wick rotate to Lorentzian signature by $\tau=i t$,
\ie
\label{FLRW}
ds^2 &= -dt^2 + 4\sin^2 t{d\rho^2+\rho^2 d\theta^2\over (1-\rho^2)^2},
\fe
and the $\rho = 1$ surface becomes the horizon of the FLRW universe.

To understand the causal structure of the full geometry, it is simplest to take the two light conical defects as created by ``heavy particles'' (whose mass is of order Planck scale but parametrically small) to avoid strong back reaction.  We propose that the Penrose diagram for the full geometry is as shown in Figure~\ref{Fig:Penrose}.  Patch I and patch II describe vacuum AdS, where two particles come in from $\theta = 0, \pi$ and collide at $t = 0$.  In patch III, the geometry 
is an FLRW universe \eqref{FLRW}
with an identification $\theta \sim \theta + 2\pi (1-2\eta)$, where $4\pi\eta$ is the deficit angle of the conical defect located at $\rho = 0$.\footnote{A small perturbation in the FLRW patch, say by some matter field, generates a big ``crunch'' in the future, where time effectively ends \cite{Hertog:aa,Hertog:ab,Kumar:2015aa}.  In Figure~\ref{Fig:Penrose}, this can be represented by shrinking the future dashed line at $t = \pi/2$ to a point.
}


There seems to be a connection between universality and the existence of a horizon.  The Cardy formula applies in the regime where the bulk thermodynamics is dominated by BTZ black holes \cite{Hartman:2014oaa}.  Recall from Section~\ref{BulkAction} that the universality of the OPE coefficients in the CFT only holds when a triangle inequality for the deficit angles is violated, which creates a horizon in the Lorentzian bulk geometry.


%




We leave for future work the gravity interpretation of the semiclassical OPE coefficients that involve operators with scaling dimensions above $c \over 12$.  Such operators correspond to BTZ microstates.  In the Lorentzian signature, these OPE coefficients could be related to the process of two conical defects merging into a BTZ blackhole, or two BTZ black holes merging into a larger BTZ black hole.  The multi-boundary wormhole geometries described in \cite{Krasnov:2000zq,Balasubramanian:2014hda} might play a role.

We end with a comparison of the semiclassical limit considered in this paper, $c\to\infty$ holding $\Delta/c$ fixed, with the more conventional limit in AdS/CFT, $c\to\infty$ with $\Delta$ fixed.\footnote{We thank the JHEP referee for suggesting this clarification.
} In bulk perturbation theory, the logarithm of the OPE coefficient, ${\mathcal P}(\Delta, c)$, takes the following expansion form,\footnote{For simplicity, we denote the dimensions $\Delta_1$, $\Delta_2$ and $\Delta_3$ collectively as $\Delta$.}
\ie
{\mathcal P}(\Delta, c) = {\mathcal P}_0(\Delta) +{ {\mathcal P}_1(\Delta)\over c} + { {\mathcal P}_2(\Delta)\over c^2} + \cdots.
\fe
The function ${\mathcal P}_0(\Delta)$ was computed by a tree-level Witten diagram in \cite{Freedman:aa}, and the functions ${\mathcal P}_n(\Delta)$ for $n \ge 1$ correspond to loop Witten diagrams. If the functions  ${\mathcal P}_n(\Delta)$ in the large $\Delta$ limit scale at most as
\ie\label{LargeDeltaScaling}
{\mathcal P}_n(\Delta)=p_n \Delta^{n+1} + {\cal O}(\Delta^{n}),
\fe
then the function ${\mathcal P}(\Delta, c)$ in the semiclassical limit can, in principle, be recovered by the partial resummation
\ie
 \lim_{c\to\infty, \text{ fixed $\Delta / c$}}{1\over c}{\mathcal P}(\Delta, c) = \sum_{n=0}^\infty p_n \left(\Delta\over c\right)^{n+1}.
\fe
Indeed, the tree-level Witten diagram satisfies the scaling condition \eqref{LargeDeltaScaling}, and the $p_0$ coincides with the heavy particle limit ($\Delta/c\to 0$) of the semiclassical OPE coefficient \eqref{WorldAction}. It would be interesting to examine the large $\Delta$ scaling of the loop Witten diagrams, and compare the coefficients $p_n$ with the $\Delta/c$ expansion of the semiclassical OPE coefficient.

\section*{Acknowledgments}

We are grateful to Tarek Anous, Ethan Dyer, Daniel Harlow, Alexander Maloney, Jonathan Maltz, Gim Seng Ng, Sinha Ritam, Simon F. Ross, Tomonori Ugajin and Xi Yin for insightful discussions, and to Tarek Anous and Gim Seng Ng for comments on the first draft of this paper.  We would like to thank the Tata Institute for Fundamental Research for support during the course of this work, and especially to the hospitality of Shiraz Minwalla.  C.M.C. is supported by BCTP Funding 39862-13070-40-PHBCTP.  Y.H.L. is supported by the Fundamental Laws Initiative Fund at Harvard University.

\appendix

\section{Special functions and their semiclassical limit}
\label{Sec:Special}

This appendix defines the special functions appearing in the fusion transformation and the DOZZ formula, and computes their semiclassical expansions.

The Barnes double gamma function $\Gamma_2(x|\omega_1,\omega_2)$ is defined as
\ie
\log\Gamma_2(x|\omega_1,\omega_2)={\partial\over \partial t}\sum^\infty_{n_1,n_2=0}(x+n_1\omega_1+n_2\omega_2)^{-t}\Big|_{t=0},
\fe
from which we define the special functions $\Gamma_b$, $S_b$, and $\Upsilon_b$,
\ie
\Gamma_b(x)={\Gamma_2(x|b,b^{-1})\over \Gamma_2(Q/2|b,b^{-1})},
\quad
S_b(x)={\Gamma_b(x)\over \Gamma_b(Q-x)},
\quad
\Upsilon_b(x)={1\over \Gamma_b(x)\Gamma_b(Q-x)}.
\fe
$\Gamma_b$ is a meromorphic function of $x$ and has poles at $x=-mb-n/b$ for non-negative integers $m$ and $n$, and satisfies the recursion relation
\ie
\label{GammaPeriodicity}
\Gamma_b(x+b) = {\sqrt{2\pi} b^{bx - 1/2} \over \Gamma(bx)} \Gamma_b(x), \quad \Gamma_b(x+1/b) = {\sqrt{2\pi} (1/b)^{x/b - 1/2} \over \Gamma(x/b)} \Gamma_b(x).
\fe

We are interested in the limit of $b \to 0$ with $bx$ fixed.  Let us define
\ie
\Lambda(y) \equiv b^2 \log \Gamma_b(y/b),
\fe
so that the recursion relation becomes a first order differential equation
\ie
\Lambda'(y) = \log\sqrt{2\pi} + (y-1/2) \log b - \log \Gamma(y) + {\cal O}(b^2).
\fe
When $y \not\in (-\infty, 0)$, the solution to this differential equation gives the semiclassical limit of the special functions\footnote{For both the prefactor $P_b$ and the contour integrand $T_b$ in the fusion kernel, the $\log b$ terms all cancel, and the $\log \sqrt{2\pi}$ terms combine into a constant that is independent of the $\eta$'s.  So loosely speaking, the semiclassical limit of the special functions $\Gamma_b, S_b, \Upsilon_b$ are $G, H, F$.
}
\ie
\label{SemiSpecial}
& b^2 \log\Gamma_b(y/b) = G(y) + (y - 1/2) \log \sqrt{2\pi} + {(y - 1/2)^2 \over 2} \log b 
+ {\cal O}(b^2),
\\
& b^2 \log S_b(y/b) = H(y) + (2y - 1) \log\sqrt{2\pi} 
+ {\cal O}(b^2),
\\
& b^2 \log \Upsilon_b(y/b) = F(y) - (y - 1/2)^2 \log b
+ {\cal O}(b^2),
\fe
where $G, H, F$ are defined as
\ie\label{defG}
G(y) &\equiv - \int_{1/2}^y \log\Gamma(z) dz, ~~ H(y) \equiv G(y) - G(1-y), ~~ F(y) \equiv - G(y) - G(1-y).
\fe
The function $G$ (also $\log\Gamma$) has a branch cut on the negative real line $(-\infty, 0)$,  and the imaginary part of of the integral in \eqref{defG} is ambiguous up to shifts of $2\pi N y$ where $N$ is an integer labeling the sheet.  Since $\Gamma_b(x)$ is real and positive for $x \in \bR_{\geq0}$, we fix this ambiguity by demanding that $G(y)$ is real for $y \in \bR_{\geq0}$.  With this definition, the special functions obey the recursion relations
\ie
\label{GIHPeriods}
G(y) &= G(y+1) + G(0) - G(1) - y + \begin{cases}
y \log y & \text{Re}\,y \geq 0
\\
y \log (-y) + \text{sgn}(\text{Im}\,y) i \pi y & \text{Re}\,y < 0
\end{cases}
\\
H(y) &= H(y+1) + 2G(0) - 2G(1) +  \text{sgn}(\text{Im}\,y) i \pi y
\\
F(y) &= F(1-y) = F(y+1) +  2 y + \begin{cases} - 2 y \log y + \text{sgn}(\text{Im}\,y) i \pi y & \text{Re} \, y \geq 0
\\
- 2 y \log (-y) - \text{sgn}(\text{Im}\,y) i \pi y & \text{Re}\,y < 0.
\end{cases}
\fe
Note that
\ie
\log\sqrt{2\pi} = G(0) - G(1), \quad G(2) = 1 + 2G(0) - G(1).
\fe

We comment on the origin of the branch cut.  For finite $b$, the function $\Gamma_b(x)$ is meromorphic when $x$ is away from the array of poles that lie on the negative real axis.  Along the negative real axis, $\Gamma_b(x)$ changes sign whenever $x$ crosses a pole, which means that $\log\Gamma_b(x)$ acquires an additional imaginary part $- i \pi$ when $x$ is above the real axis, and $i \pi$ when below.  In the semiclassical  $b \to 0$ limit, the poles become densely populated on the negative real axis, and create a branch cut across which the imaginary part is discontinuous.

When $y \in (-\infty, 0)$, we can make use of the second recursion realtion \eqref{GammaPeriodicity} of the $\Gamma_b$ function to define $\Gamma_b(y/b)$ in terms of $\Gamma_b((y+1)/b)$.  To take the semiclassical limit, we need the asymptotics of the $\Gamma$ function,
\ie
\Gamma(y/b^2) = { 1 \over e^{i \pi y/b^2} - e^{- i \pi y/b^2} } \exp\left[{1\over b^2} (y \log(-y/b^2) - y) + {\cal O}(\log b) \right], \quad \text{Re}\,y < 0.
\fe
We do not need this expression when we take the semiclassical limit of the fusion transformation in Section~\ref{Sec:Fusion}, since the arguments there have small imaginary regulators.

%

\section{Semiclassical Virasoro blocks}
\label{SemiclassicalCB}

In the limit of large central charge $c$ while taking the operator weights $h_i$ to scale with $c$ (fixed $m_i = {h_i \over c}$), the Virasoro block exponentiates as \eqref{BlockClassExp}  \cite{Zamolodchikov:426555}, which means that the limit
\ie
\label{SemiBlockDef}
f\Big({h_{ext}\over c},{h\over c} \Big| x\Big) \equiv \lim_{c\to\infty} {6 \over c} \log {\cal F}(h_{ext}, h, c | x)
\fe
exists.  The function $f(h_{ext}/c,h/c|x)$ is referred to as the semiclassical Virasoro block and can be computed order by order in an $x$-expansion. To third order in the $x$-expansion,
\ie
\label{SecondOrderExpansion}
{c\over 6}f\Big({h_{ext}\over c},{h\over c} \Big| x\Big) &= (2h_{ext} - h) \log x - {h x \over 2} - { 3h + 26h^2 + 16h_{ext}h + 32h_{ext}^2 \over 16 (1+8h)} x^2
\\
& \hspace{1in} -\frac{46 h^2+48 h h_{ext}+5 h+96 h_{ext}^2}{384 h+48}x^3 + {\cal O}(x^4).
\fe

The radius of convergence of any Virasoro block as a function of $x$ is unity.  After factoring out a power of $x$, the only potential poles are at $1$ and $\infty$.  However, this does not guarantee that the semiclassical Virasoro block has the same radius of convergence.  Due to the logarithm in the definition \eqref{SemiBlockDef}, its radius of convergence is determined not only by the poles but also the zeros in the Virasoro block $\cal F$.  Let us make two comments:
\begin{enumerate}
\item  When Virasoro blocks are computed numerically, expanding in the nome $q(x)$ instead of $x$ gives a much faster rate of convergence.  The map from $x$ to $q$ maps the entire complex plane to a region within the unit disk, and the interval $(0, 1)$ to $(0, 1)$ itself.  Since all the zeros of a Virasoro block is mapped to inside the unit disk, the radius of convergence of the $q$-expansion will typically be worse than that of the $x$-expansion.
\item  For unitary values of central charges and weights, the Virasoro block after factoring out a conformal factor has an $q$-expansion with non-negative coefficients, because these coefficients can be regarded as the norm of a state in the Hilbert space of quantizing the CFT in a pillow geometry \cite{Maldacena:2015iua}.  The non-negativity implies that there is no zero in the interval $q \in (0, 1)$.  Hence for any given Virasoro block, we can always find a holomorphic variable transformation that maps $(0, 1)$ to $(0, 1)$ but moves all the zeros outside the unit disk.  When expanded in this new variable, the radius of convergence of the Virasoro block is unity.  In the semiclassical limit, if the set of zeros do not become arbitrarily close to the interval $(0, 1)$, then likewise there exists a variable transformation such that the semiclassical Virasoro block also has unit radius of convergence.
\end{enumerate}

Relatedly, the semiclassical Virasoro block is the leading term in the asymptotic $1/c$ expansion, so there might exist non-perturbative error terms that are exponentially suppressed when $x$ is small but become large otherwise.\footnote{This effect was demonstrated in the heavy-light limit by \cite{Fitzpatrick:2016ta,Kaplan:2016ta,Fitzpatrick:2016aa}.
}
To illustrate this, let us say that the Virasoro block has a semiclassical expansion of the form
\ie\label{AsymVir}
{\cal F}(x) = \exp\left[ - {c\over 6} f_1(x)+  {\cal O}(c^0)\right] + \exp\left[- {c\over 6} f_2(x)+  {\cal O}(c^0) \right] + \cdots,
\fe
and we assume an ordering $\text{Re}\,f_1(x) < \text{Re}\,f_2(x) < \cdots$ that is valid in a neighborhood near $x=0$.  When we compute the semiclassical Virasoro block as a series in $x$, we are implicitly assuming that $x$ is inside this neighborhood, hence what we get is $f_1(x)$.  Outside this neighborhood, the Virasoro block may undergo a ``phase transition", i.e., some $f_i(x)$ may have a smaller real part than $f_1(x)$, and the semiclassical approximation by $f_1(x)$ completely breaks down.

The good news is that the fusion transformation gives us a handle on testing the radius of convergence of the semiclassical Virasoro block and the (non)existence of a phase transition in the region $x \in (0,1)$.  Let us focus on the vacuum block. In Section \ref{Sec:Fusion}, we evaluated the semiclassical limit of the fusion transformation. By the semiclassical fusion kernel \eqref{SemiF} and the steepest descent equations \eqref{SPE3} and \eqref{SPE1}, the vacuum block can be written as
\ie
{\cal F}(h_{\A_{ext}}, 0, c | x) &\approx {\bf F}^{(c)}_{0 \A_t}
[ \A_{ext}]
 {\cal F}(h_{\A_{ext}}, h_{\A_t}, c | 1-x)
 \\
 &\approx \exp\left[ \log {\bf F}^{(c)}_{0 \A_t}[ \A_{ext}] - {c \over 6} f\Big({h_{\A_{ext}}\over c}, {h_{\A_t}\over c} \Big| 1-x\Big) \right],
\fe
where ${\A_t}$ is the critical point of the steepest descent approximation, which depends on $x$.  The function $f(h_{\A_{ext}}/c,h/c|1-x)$ can be computed as an expansion in $1-x$.  One can check whether the $x$-expansion works at the desired value of $x$ by comparing the two sides.

Let us end with an example where we know that the radius of convergence of the $x$-expansion is one: when $m_{ext}, m \ll 1$, the semiclassical Virasoro block to linear order in $m_{ext}, m$ has an exact expression
\ie\label{LinearSemiBlock}
{c \over 6} f\Big({h_{ext}\over c},{h\over c} \Big| x\Big) = (2h_{ext}-h) \log\left[{4(2-x-2\sqrt{1-x})\over x}\right] - 4h_{ext}\log\left[2-2\sqrt{1-x}\over x\right],
\fe
that is obtained from a bulk worldline computation \cite{Chang:2015aa}.

\section{On-shell Liouville action}
\label{OSLiouville}

In this section, we review the solution to the Liouville equation
\ie
\partial\bar\partial\varphi=2\pi \mu b^2 e^\varphi
\fe
with boundary conditions
\ie
\varphi(z,\bar z) \to
\begin{cases}
-2\log|z|^2 & z \to \infty
\\
-2\eta_i\log|z-z_i|^2 & z\to z_i,
\end{cases}
\fe
and evaluate the on-shell classical Liouville action
\ie
\label{SL}
S_L=\int_{\Gamma} d^2 z\,{1\over 4\pi}\left(\partial\varphi\bar\partial\varphi +4\pi \mu b^2 e^{\varphi}\right) +\left(\varphi_\infty+2\log R\right)-{\textstyle\sum_i}\left(\eta_i\varphi_i+2\eta_i^2\log\epsilon_i\right).
\fe
We closely follow the calculation in \cite{Zamolodchikov:1995aa,Harlow:2011aa}, but instead of a positive cosmological constant
 $\mu$, we consider a negative one.\footnote{The Ricci curvature of the metric $ds^2=e^{\varphi}dzd\bar z$ is ${\cal R}=-8\pi\mu b^2$, so negative $\mu$ implies positive curvature.} 
The $\eta_i$ appearing in the boundary condition have the interpretation of conical defects of deficit angle $4\pi\eta_i$.  They are assumed to be in the range $0 \le \eta_i \le {1\over 2}$ so that the deficit angles are at most $2\pi$.

The Liouville equation can be solved by the ansatz
\ie\label{Anzf}
e^\varphi={1\over \pi|\mu|b^2f(z,\bar z)^2},
\fe
where the function $f(z,\bar z)$ must satisfy the differential equation
\ie\label{LiouvilleEqF}
&\partial\bar\partial f={1\over f}(\partial f\bar\partial f+1)
\fe
and the boundary conditions
\ie
f(z,\bar z) \propto \begin{cases} |z|^2 & z \to \infty
\\
|z-z_i|^{2\eta_i} & z\to z_i.
\end{cases}
\fe

To proceed, let us define
\ie
W= - {\partial^2 f \over f}, \quad \widetilde W= - {\bar \partial^2 f \over f}.
\fe
By the equation of motion \eqref{LiouvilleEqF}, one can show that $W$ is holomorphic and $\widetilde W$ is anti-holomorphic.  The boundary conditions on $f(z,\bar z)$ then uniquely fix $W(z)$ to be
\ie
&W(z) = {1\over (z-z_1)(z-z_2)(z-z_3)} \left[{\eta_1(1-\eta_1)z_{12}z_{13}\over z-z_1} + (\text{2 permutations})
\right],
\fe
where $z_{ij} \equiv z_i - z_j$, and $\widetilde W(\bar z)$ is given by the replacements $z \to \bar z$ and $z_i \to \bar z_i$.

Now $f(z,\bar z)$ satisfies a holomorphic and an anti-holomorphic differential equation
\ie\label{holEqf}
\partial^2 f + W(z) f = 0, \quad \bar\partial^2 f + \widetilde W(\bar z) f = 0.
\fe 
Each of these equations takes the form of Riemann's hypergeometric differential equation.  The solution is given by
\ie
\label{Solf}
&f(z,\bar z)=a_1 u(z)\overline{u( z)}- a_2v(z)\overline{v( z)},
\fe
where
\ie
&u(z)=(z-z_2)x^{\eta_1}(1-x)^{\eta_3}{}_2F_1(\eta_1+\eta_3-\eta_2,{\textstyle\sum_i\eta_i}-1,2\eta_1,x),
\\
&v(z)=(z-z_2)x^{1-\eta_1}(1-x)^{1-\eta_3}{}_2F_1(1+\eta_2-\eta_1-\eta_3,2-{\textstyle\sum_i\eta_i},2-2\eta_1,x),
\fe
and
\ie
&x = {(z-z_1)z_{32} \over (z-z_2)z_{31} }.
\fe

We are left with two undetermined coefficients $a_1$ and $a_2$.  Plugging the solution \eqref{Solf} back into the equation of motion \eqref{LiouvilleEqF} gives
a relation between the two coefficients
\ie\label{a1a2}
a_1a_2=-{|z_{13}|^2\over |z_{12}|^2|z_{23}|^2(1-2\eta_1)^2}.
\fe
A second condition comes from demanding the single-valuedness of the function $f(z,\bar z)$, in particular near $z=z_3$. 
The final solution is
\ie\label{Coelf}
&a_1^2={|z_{13}|^2\over |z_{12}|^2|z_{23}|^2}{\gamma({\textstyle\sum_i\eta_i})\gamma(\eta_1+\eta_2-\eta_3)\gamma(\eta_1+\eta_3-\eta_2)\over (1-{\textstyle\sum_i\eta_i})^2\gamma^2(2\eta_1)\gamma(\eta_2+\eta_3-\eta_1)},
\\
&a_2^2={|z_{13}|^2\over |z_{12}|^2|z_{23}|^2}{ (1-{\textstyle\sum_i\eta_i})^2 \gamma^2(2\eta_1)\gamma(\eta_2+\eta_3-\eta_1)\over (1-2\eta_1)^4\gamma({\textstyle\sum_i\eta_i})\gamma(\eta_1+\eta_2-\eta_3)\gamma(\eta_1+\eta_3-\eta_2) },
\fe
where $\gamma(y) \equiv {\Gamma(y) \over \Gamma(1-y)}$.
If the triangle inequalities for the three $\eta_i$ and also $\eta_1+\eta_2+\eta_3 \leq 1$ are satisfied, $a_1$ and $a_2$ are real.  In this case, since $a_1$ and $a_2$ have opposite signs due to \eqref{a1a2}, 
We can choose $a_1 > 0$ and $a_2 < 0$, so that $e^\varphi$ as given by \eqref{Anzf} and \eqref{Solf} is positive and has poles only at $z_1$, $z_2$ and $z_3$. 
If one of the inequalities is violated, $a_1$, $a_2$ are pure imaginary.  Not only is $e^\varphi$ negative, but now it is possible for $e^\varphi$ to diverge at points other than $z_1$, $z_2$ and $z_3$.  

%
%
%

Now we evaluate the classical Liouville action \eqref{LiouvilleA} on the solution we just found.  We adopt the same trick as in \cite{Zamolodchikov:1995aa,Harlow:2011aa}, which is to first consider the derivative of the classical action $S_L$ with respect to $\eta_i$.  When evaluated on a classical solution, $S_L$ depends both explicitly on $\eta_i$ through the boundary terms in the Liouville action \eqref{SL} and implicitly on $\eta_i$ through the classical solution,
\ie\label{pSpEta}
{d S_L\over d \eta_i}={\partial S_L\over \partial \eta_i}+{\delta S_L\over \delta\varphi} {\partial\varphi\over\partial\eta_i}.
\fe
The second term vanishes on-shell, hence the derivative only receives contribution from the boundary terms,
\ie
{d S_L\over d \eta_i} = - \varphi_i + 4 \eta_i \log \epsilon_i.
\fe
Expanding our solution around $z = z_i$, we find
\ie\label{varphiExZ}
\varphi(z,\bar z) &\to -2 \eta_i\log|z-z_i|^2 + C_i,
\fe 
where
\ie\label{LiouAC}
C_1 &= 2\pi i N - \log \pi|\m|b^2 -(1-2\eta_1)\log{|z_{12}|^2|z_{13}|^2\over |z_{23}|^2}
\\
&\hspace{1.5in} -\log{\gamma({\textstyle\sum_i\eta_i})\gamma(\eta_1+\eta_2-\eta_3)\gamma(\eta_1+\eta_3-\eta_2)\over (1-{\textstyle\sum_i\eta_i})^2\gamma^2(2\eta_1)\gamma(\eta_2+\eta_3-\eta_1)},
\fe
and $C_2$ and $C_3$ are given by cyclically permuting the $\eta_i$.  Here $N \in \bZ$ labels the ambiguity of shifting any classical solution $\varphi$ by $2\pi i$.  The logarithmic divergence cancels with the regulator, and we end up with
\ie
{dS_L \over d\eta_i} = - C_i.
\fe
It is then straightforward to integrate with respect to $d\eta_i$ and obtain the action itself
\ie
S_L &= ({\textstyle\sum_i\eta_i} - 1)\log\pi|\m|b^2+2(1-{\textstyle\sum_i\eta_i})\left[\log(1-{\textstyle\sum_i\eta_i}) -1 + \pi i N \right] + F({\textstyle\sum_i\eta_i}) 
\\
& \hspace{-.5in} - F(0) + \left\{ (\delta_2+\delta_3-\delta_1)\log|z_{23}|^2+F(\eta_2+\eta_3-\eta_1) - F(2\eta_1) + (\text{2 permutations}) \right\},
\fe
where
\ie
F(y) \equiv \int_{1/2}^y \gamma(z) dz, \quad \delta_i \equiv \eta_i (1-\eta_i).
\fe
As in \cite{Zamolodchikov:1995aa}, the integration constant can be fixed by matching with the special case $\eta_1+\eta_2+\eta_3=1$, where the known answer is
\ie
S_L=\sum_{i<j}2\eta_i\eta_j\log|x_i-x_j|^2.
\fe

\section{Violation of triangle inequality}
\label{Sec:violation}

In this appendix, we discuss issues when the triangle inequality is violated, 
and show that the properly regularized gravity action is still real.

When one of the triangle inequalities is violated, $\eta_1>\eta_2+\eta_3$, the Liouville field $\varphi$ becomes multivalued and has branch cuts. The imaginary part of $\varphi$ is piecewise constant and jumps across the branch cuts. The function $f(z,\bar z)$, related to $\varphi$ by \eqref{Anzf}, is still single-valued, but vanishes on the branch cut.  
We check that for the explicit solution \eqref{Solf}, the branch cut is a loop that encloses the point $z_1$ but not $z_2$ and $z_3$.  The Liouville action \eqref{SL} can be written as
\ie
\label{SL}
S_L
&=\int_{\Gamma} d^2 z\,{1\over \pi}\left({\partial f\bar\partial f+1\over f^2}\right) +\left(\varphi_\infty+2\log R\right)-{\textstyle\sum_i}\left(\eta_i\varphi_i+2\eta_i^2\log\epsilon_i\right).
\fe
Let us denote the imaginary part of $\varphi$ by $\theta$.  The first term is real and independent of the imaginary part of $\varphi$.  The other terms give a contribution 
\ie\label{LAI}
i(\theta_\infty+{\textstyle\sum_i}\eta_i\theta_i),
\fe
where $\theta_\infty, \, \theta_i$ are the imaginary part of $\varphi$ at $\infty,\, z_i$.  By inspecting the behavior of $\varphi$ at $\infty, \, z_i$ given in \eqref{varphiExZ} and \eqref{LiouAC}, we find\footnote{The analytic continuation of \eqref{ConicalOPE}, whose imaginary part is given in \eqref{ImaginaryPart}, does not contain the contribution from $\theta_\infty$.
}
\ie
\theta_1=\pm \pi, \quad \theta_2=\theta_3=\theta_\infty=\mp\pi.
\fe

Now let us consider the gravity action. The cutoff surface is modified to
\ie
r_{max}=1-\epsilon |e^{\varphi\over 2}|=1-\epsilon e^{{\varphi\over 2} - i {\theta\over 2}}.
\fe
Since $|e^{\varphi\over 2}|\propto f^{-1}$ diverges on the branch cut, for the cutoff surface to be well-defined, we need to regularize by modifying the integration domain of the $z$-integral to $\Gamma \setminus \cW$,
where $\cal W$ is a neighborhood of the branch cut. The regularized volume and area are given by
\ie
&V_\epsilon=\int_{\Gamma\setminus\cW} d^2 z\left[{e^{i\theta}\over 2\epsilon^2}-{e^{{\varphi\over 2}+{i\theta\over 2}}\over 2\epsilon}+{1\over 8}e^{\varphi}\left(1+2\varphi - 2i\theta+4\log{\epsilon\over 2}\right)\right]+{\cal O}(\epsilon), 
\\
&A_\epsilon=\int_{\Gamma\setminus\cW} d^2 z\left[{e^{i\theta}\over \epsilon^2}-{e^{{\varphi\over 2}+{i\theta\over 2}}\over \epsilon}+{1\over 8}\left(-2 e^{\varphi}+4\partial(\varphi-i\theta)\bar\partial(\varphi-i\theta)\right)\right]+{\cal O}(\epsilon).
\fe
The regularized gravity action is
\ie
&V_\epsilon-{1\over 2}A_\epsilon
\\
&=\int_{\Gamma\setminus\cW} d^2 z\left[{1\over 4}\left(\partial\varphi\bar\partial\varphi - e^{\varphi}\right) -\left(1+\log{\epsilon\over 2}\right)\partial \bar\partial \varphi- {1\over 2}\bar\partial(\varphi\partial\varphi)+{1\over 4}i(\partial(\theta\bar\partial\varphi)+\bar\partial(\theta\partial\varphi))+{1\over 4}\partial\theta\bar\partial\theta\right]
\\
&=\pi S_L\big|_{\pi\mu b^2=-{1\over 4}}+2\pi\left(1-\log 2 + \log\epsilon\right) (1-{\textstyle\sum_i}\eta_i)-2\pi\log R
+ 2\pi{\textstyle\sum_i}\eta_i^2\log\epsilon_i- i\pi(\theta_\infty+ {\textstyle\sum_i}\eta_i\theta_i).
\fe
On the second line, the last term is only nonzero inside $\cal W$, and hence does not contribute;
the third and forth terms can potentially produce boundary terms on $\partial\cal W$, but their contributions cancel.  In the final expression, the imaginary last term cancels the imaginary part of the first term, which is $\pi$ times \eqref{LAI}.

%

\section{Semiclassical Liouville CFT}
\label{Sec:Liouville}

The Liouville CFT of central charge $c = 1 + 6Q^2$ ($Q = b+1/b$) and cosmological constant $\mu$ has a continuous spectrum of scalar primaries, which are exponential operators $e^{\A\phi}$ with $\A \in {Q \over 2} + i\bR_{\geq0}$.  We define the semiclassical limit to be the limit of $b \to 0$ with fixed $\eta \equiv \A b$.
In this limit, the spectrum of primaries in Liouville theory are parameterized by $\eta = {1\over 2} - \sqrt{{1\over 4}-{6h \over c}} \in {1 \over 2} + i \bR_{\geq0}$, where $h$ is the weight.  We may also consider non-normalizable operators of weight $h < {c\over24}$ corresponding to $\eta \in [0, {1 \over 2}]$, though they do not lie in the Hilbert space of the Liouville CFT.

The exponential operators are normalized by the reflection amplitude
\ie
S(\A) = - (\pi\mu b^2)^{(Q - 2\A) / b} { \Gamma(1-{(Q - 2\A) / b}) \Gamma(1-{(Q - 2\A) b } ) \over \Gamma(1+{(Q - 2\A) / b}) \Gamma(1+{(Q - 2\A) b} ) },
\fe
whose semiclassical limit is
\ie
\lim_{b \to 0} {b^2} \log S(\eta/b) &= \left[ (1-2\eta) ( 2 + \log(\pi\mu b^2) - 2 \log(1-2\eta) ) + {\rm sgn}({\rm Im}\,\eta) i\pi(2\eta-1) \right].
\fe
The three-point function coefficients are given by the DOZZ formula
\ie
{C}(\A_1, \A_2, \A_3) &= \left[ \pi\mu \gamma(b^2) b^{2-2b^2} \right]^{(Q - \sum_i \A_i)/b} 
\\
& \hspace{-.5in} \times { \Upsilon'_b(0) \over \Upsilon_b({\textstyle\sum_i \A_i} - Q) } \left[ { \Upsilon_b(2\A_1) \over \Upsilon_b(\A_2 + \A_3 - \A_1) } \times (\text{2 permutations}) \right],
\fe
whose semiclassical limit is
\ie
\lim_{b \to 0} {b^2}\log {C}(\eta_1/b, \eta_2/b, \eta_3/b) 
& = - \Big[
({\textstyle\sum_i \eta_i} - 1) \log(\pi\mu b^2) - F(0) + F({\textstyle\sum_i \eta_i} - 1)
\\
&\hspace{0in} + \{ F(\eta_2 + \eta_3 - \eta_1) - \sum_i F(2\eta_1) + (\text{2 permutations}) \} \Big].\fe
Recall from Appendix~\ref{Sec:Special} that $F(y) \equiv \int^y_{1\over 2}\log\gamma(x) dx$ is the semiclassical limit of the special function $\Upsilon_b$.

We point out a small observation: when $h_{ext}, h_t > {c\over24}$, the semiclassical fusion kernel \eqref{SemiF} or \eqref{FKR2} can be written in terms of the reflection amplitude $S$ and the DOZZ three-point function ${C}$ of the Liouville CFT,
and the holomorphic Cardy formula
\ie
\rho(h) = \exp\left[ {2 \pi} \sqrt{{c\over6} ( h - {c \over 24} ) } \right]
\fe
as
\ie
\hspace{-.1in} {\bf F}^{(6/b^2)}_{0,\A_t} [\A_{ext}]& 
= \exp\left[ {1 + \log\sqrt{\pi\mu b^2} \over b^2} + {\cal O}(\log b) \right] { \rho(h_{\A_t}) {C}(\A_{ext}, \A_{ext}, \A_t) \over \rho(h_{\A_{ext}}) S(\A_{ext}) \sqrt{ \rho(h_{\A_t}) S(\A_t)} }.
\fe
However, the Liouville CFT does not have a normalizable vacuum state, while the validity of interpreting the fusion kernel as the semiclassical OPE coefficient hinges crucially on the existence of a normalizable vacuum state.

\bibliography{refs} 
\bibliographystyle{JHEP}

\end{document}